\documentclass[aoas]{imsart}
\usepackage{fix-cm}

\RequirePackage{amsthm, amsmath, amsfonts, amssymb}
\RequirePackage{natbib}
\RequirePackage{enumerate}
\RequirePackage[colorlinks, citecolor=blue, urlcolor=blue]{hyperref}
\RequirePackage{graphicx}
\RequirePackage{caption, subcaption}
\usepackage{thmtools}
\usepackage{tikz}
\usetikzlibrary{shapes, decorations.pathreplacing, calc, arrows.meta, fit, positioning, automata}
\let\singlespacing\relax
\let\doublespacing\relax
\usepackage{setspace}

\startlocaldefs
\theoremstyle{plain}

\newtheorem{assumption}{Assumption}
\newtheorem{assumptionp}{Assumption}  
\newtheorem{theorem}{Theorem}[section]

\theoremstyle{remark}

\newcommand{\btheta}{\boldsymbol{\theta}}

\newcommand{\bbeta}{ \boldsymbol{\beta}}

\newcommand{\bdelta}{ \boldsymbol{\delta}}

\newcommand{\bomega}{ \mbox{\boldmath $\omega$}}

\newcommand{\bc}{ \mbox{\bf c}}

\newcommand{\bX}{\mathbf{X}}

\newcommand{\bs}{\mathbf{s}}

\newcommand{\bU}{ \mbox{\bf U}}

\newcommand{\bV}{ \mbox{\bf V}}

\newcommand{\bD}{ \mbox{\bf D}}

\newcommand{\bw}{ \mbox{\bf w}}

\newcommand{\bR}{ \mbox{\bf R}}

\newcommand{\calR}{{\cal R}}

\newcommand{\calG}{{\cal G}}
\newcommand{\calD}{{\cal D}}
\newcommand{\calC}{{\cal C}}
\newcommand{\calS}{{\cal S}}

\newcommand{\calA}{{\cal A}}

\newcommand{\calX}{{\cal X}}

\newcommand{\calU}{{\cal U}}
\newcommand{\calV}{{\cal V}}
\newcommand{\calM}{{\cal M}}
\newcommand{\calW}{{\cal W}}
\newcommand{\calY}{{\cal Y}}

\newcommand{\beq}{ \begin{equation}}
\newcommand{\eeq}{ \end{equation}}
\newcommand{\beqn}{ \begin{eqnarray}}
\newcommand{\eeqn}{ \end{eqnarray}}

\usepackage{caption}
\renewcommand{\arraystretch}{1.5}

\endlocaldefs

\begin{document}

\begin{frontmatter}
\title{Spatial causal inference in the presence of preferential sampling to study the impacts of marine protected areas}
\runtitle{Spatial causal inference in the presence of preferential sampling}

\begin{aug}
\author[A]{\fnms{Dongjae}~\snm{Son} 
\ead[label=e1]{dson@ncsu.edu}},
\author[A]{\fnms{Brian J.}~\snm{Reich}\ead[label=e2]{bjreich@ncsu.edu}\orcid{0000-0000-0000-0000}}
\author[A]{\fnms{Erin M.}~\snm{Schliep}\ead[label=e3]{emschliep@ncsu.edu}}
\author[A]{\fnms{Shu}~\snm{Yang}\ead[label=e4]{syang24@ncsu.edu}}
\and
\author[B]{\fnms{David A.}~\snm{Gill}\ead[label=e5]{david.gill@duke.edu}}
\address[A]{Department of Statistics,
North Carolina State University\printead[presep={ ,\ }]{e1,e2,e3,e4}}

\address[B]{Division of Marine Science and Conservation, Duke University
\printead[presep={,\ }]{e5}}
\end{aug}

\begin{abstract}
Marine Protected Areas (MPAs) have been established globally to conserve marine resources. Given their maintenance costs and impact on commercial fishing, it is critical to evaluate their effectiveness to support future conservation. In this paper, we use data collected from the Australian coast to estimate the effect of MPAs on biodiversity. Environmental studies such as these are often observational, and processes of interest exhibit spatial dependence, which presents challenges in estimating the causal effects. Spatial data can also be subject to preferential sampling, where the sampling locations are related to the policy and the response variable, further complicating inference and prediction. To address these challenges, we propose a spatial causal inference method that simultaneously accounts for unmeasured spatial confounders in both the sampling process and the treatment allocation. We prove the identifiability of key parameters in the model and the consistency of the posterior distributions of those parameters.  We show via simulation studies that the causal effect of interest can be reliably estimated under the proposed model. The proposed method is applied to assess the effect of MPAs on fish biomass. We find evidence of preferential sampling and that properly accounting for this source of bias impacts the estimate of the causal effect.
\end{abstract}


\end{frontmatter}

\section{Introduction}\label{sec:intro}

Marine Protected Areas (MPAs) are spatially-demarcated areas of the ocean designated for the long-term conservation of marine biodiversity \citep{iucn2018applying}. Given their broad application and rapid expansion globally ($\approx$8\% of the world's oceans), research interest on MPA effectiveness has also increased over the last two decades \citep{gill2019social, unep2018ngs}. For example, \cite{edgar2014global} assessed the impact of MPAs on fish population globally, and identified certain design attributes that seem to contribute to more positive effects. \textcolor{black}{Specifically, they found that the MPAs that prohibit fishing, that are efficiently enforced, old ($> 10$ years), large ($>100\text{ km}^2$), and surrounded by deep water ($>25$ m) or expansive sand area tend to have higher fish biomass (kg per $250\text{ m}^2$).} \cite{gill2017capacity, gill2024diverse} synthesized a global database of ecological, socioeconomic, and management factors of MPAs to investigate whether MPAs are being effectively and equitably managed, and the effect of management on conservation impacts. They found that MPAs with adequate staff capacity and appropriate regulations had much greater positive impacts compared to others. \cite{desbureaux2024long} evaluated the impact of MPAs on fish productivity and local economy in coastal villages in Tanzania. \textcolor{black}{They concluded that communities near MPAs experienced a substantial improvement in local economy compared to those remote from the MPAs, while this benefit is not related to higher catches.}

Despite the large body of literature on MPA impacts, few studies employ causal inference methods to assess their effects. In a review of almost 200 MPA studies, \cite{ferraro2019causal} found that very few studies accounted for confounding spatial and temporal factors that could bias impacts (e.g. placement bias). Since prohibiting fishing activities generates greater conservation outcomes \citep{grorud2021mpa} but could also have a negative effect \textcolor{black}{on} local communities \citep{kamat2014ocean}, scientific analyses of the effect on MPAs are important. This gives rise to the need to develop a spatial causal inference framework to properly assess policy efficacy.

Environmental studies such as analyses of MPA effects often rely on observational data to examine the effect of treatments or policies on ecological phenomena, as randomized trials are challenging to apply. Therefore, causal inference methods for observational studies are required to identify and estimate causal effects \citep{imbens2015causal, hernan2010causal}. However, unlike various causal inference methods for independent data, ecological data often have spatial correlation. Spatial causal inference frameworks have recently been developed that account for spatial dependencies \citep{jarner2002estimation, davis2019addressing, schnell2020mitigating, marques2022mitigating, guan2023spectral, papadogeorgou2022causal}. These methods address unmeasured spatial confounders, which violate the key assumption in causal inference that there are no unmeasured confounders \citep{rubin1974estimating, rubin1978bayesian} for observational studies. For example, \cite{davis2019addressing} estimate propensity scores using a spatial logistic regression with conditional autoregressive priors and use them to estimate the average treatment effect by augmented inverse propensity weighting. \cite{schnell2020mitigating} propose a neighborhood adjustments method by spatial smoothing to mitigate the effect of unobserved spatial confounding. \cite{guan2023spectral} developed an approach using spectral representations to adjust for spatial confounders. \textcolor{black}{\cite{papadogeorgou2022causal} proposed a causal inference framework when the outcome and treatment follow spatiotemporal point processes.} For a more extensive review of existing spatial causal inference methods, see \cite{reich2021review}.

In addition to spatial confounding, ecological data analyses are often complicated by preferential sampling. Preferential sampling refers to the situation where the selection of sampling locations is dependent on the spatial processes of interest \citep{pati2011bayesian, diggle2010geostatistical}. For example, in our motivating MPA analysis, it may be that data are collected where they are easily accessible (e.g., close to the shoreline) where fish abundance might be low due to higher impacts from land-based activities including fishing \citep{campbell2020fishing, cinner2013global}. Preferential sampling has been shown to produce biases in parameter estimation or spatial prediction \citep{diggle2010geostatistical, pati2011bayesian, gelfand2012effect, schliep2023correcting}. \cite{diggle2010geostatistical} proposed the foundational geostatistical model that accounts for preferential sampling using a shared spatial random effect between a response model and a point process model. \cite{pati2011bayesian} proposed a Bayesian hierarchical model for which there is a theoretical guarantee that the posterior distributions of the model parameters are consistent and a proper posterior can be derived for the parameters regarding the degree of preferential sampling. \cite{gelfand2012effect} proposed a simulation-based approach to assess the effect of preferential sampling on spatial prediction. \textcolor{black}{\cite{cecconi2016preferential} proposed Bayesian hierarchical models for both continuous and discretized spatial domains with several case studies to examine the effect of preferential sampling on inference and prediction.} \cite{schliep2023correcting} proposed a weighted composite likelihood approach for estimating spatial covariance parameters under preferential sampling when the focus is on spatial prediction. \textcolor{black}{\cite{shirota2022preferential} extended the Bayesian hierarchical model used for preferential sampling to bivariate outcomes, applying possibly different specifications of point processes to each outcome. In particular, they studied how the specifications of point processes and degrees of dependence between bivariate outcomes influence inference and prediction.}

To our knowledge, there does not exist a spatial causal inference method that mitigates the effect of preferential sampling bias common in environmental studies. In this paper, we introduce a novel approach to spatial causal inference that accounts for 
preferential sampling bias. Motivated by \cite{diggle2010geostatistical} and \cite{pati2011bayesian}, we specify log-Gaussian Cox processes to model point patterns of sampling locations, which are connected to potential outcome models through shared latent spatial processes. Assuming a binary policy, two policy groups have varying degrees of preferential sampling.
By modeling the sampling intensities for each policy group, we can also calculate propensity scores using the result from superposition of Poisson processes \citep{durrett1999essentials}. In this way, we can specify a spatial causal inference model that accounts for preferential sampling in a unified hierarchical model. We present theoretical results related to the identifiability of the parameters of interest and the weak consistency of the posterior distribution.

The proposed method is applied to the MPA data presented in \cite{gill2024diverse} with a focus on Australia. The Australian marine jurisdictional area is notable for its diverse and distinctive range of species and habitats \citep{knowles2015establishing}. MPAs have been established to protect such ecological and economic resources. Across this spatial domain, we investigate the causal effect of MPAs on fish biomass by comparing MPAs and non-MPAs.

The rest of the paper is composed as follows. \textcolor{black}{In Section 2, we define the causal effects of interest and the necessary assumptions for identification, along with the proposed model for spatial causal inference that accounts for preferential sampling. In Section 3, we introduce computational details for feasibility and theoretical results, including the identification of model parameters and posterior consistency.} In Section 4, an extensive simulation study is presented to assess the performance of the proposed model. In Section 5, we apply the model to the fish biomass data and examine the effect of the MPAs. In Section 6, concluding remarks are given.

\section{Statistical methods}
\label{sec:meth}

\subsection{A general description of the modeling strategy}\label{higher-level-intro}

\textcolor{black}{Before we formally introduce the proposed framework to this problem, we explain the role of latent spatial processes as unobserved confounders, how spatial locations are preferentially sampled by these confounders, the impact of preferentially sampled locations to the estimation of causal effects, and the modeling strategy to mitigate this bias. As noted in Section~\ref{sec:intro}, sampling of spatial locations depends stochastically on underlying processes of interest in the presence of preferential sampling. This implies that the latent spatial processes are unmeasured confounders, which violates one of the standard causal inference assumptions and results in biased estimates of the causal effects. Previous literature in spatial statistics has also reported on biased parameter estimation and spatial prediction in the presence of preferential sampling \citep{diggle2010geostatistical,pati2011bayesian}. To alleviate this bias, we model the response and sampling location processes jointly through shared \emph{latent} spatial confounder processes. The details of the model are given in the following section.}

\subsection{A potential outcome approach and definition of the causal effect}\label{continuous_model}

Denote the spatial domain of interest as $\calD\subseteq\mathbb{R}^2$. There exist two policies, denoted $a=0$ and $a=1$, which in the MPA analysis represent no protection and protection policies, respectively. 
Under each policy $a\in\left\{0,1\right\}$, let us define a potential outcome $Y_a(\bs)$ at location $\bs\in\calD$. \textcolor{black}{This represents the outcome, such as fish biomass in our example, that would be observed at each location $\bs\in\calD$ for each policy group $a\in\left\{0,1\right\}$.}  Let $\bX(\bs)$ be a $p$-dimensional vector of covariates or pretreatment variables and define $\calX = \{\bX(\bs):\bs\in\calD\}$. Let $A(\bs)$ denote the policy indicator such that $A(\bs)=a$ if location $\bs$ is assigned policy $a$. The estimands of interest are the local and average causal effects. The local causal effect is
\begin{equation}
    \triangle(\bs) = \mbox{E}\left\{Y_1(\bs) - Y_0(\bs)\right\}.
    \label{local_causal_continuous}
\end{equation}
The average causal effect, referred to as the average policy effect (APE), is
\begin{equation}
    \triangle = |\calD|^{-1}\int_{\calD} \triangle(\bs) d\bs.
    \label{ate}
\end{equation}

The fundamental problem of causal inference is that at most one of the potential outcomes is observed at location $\bs$. This implies that both $Y_0(\bs)$ and $Y_1(\bs)$ are not simultaneously observable and leads us to view causal inference as a missing data problem \citep{ding2018causal}. \textcolor{black}{Identifying causal effects in settings without preferential sampling or spatial confounding requires the following assumptions:
\begin{assumption}
    Stable Unit Treatment Values (SUTVA): For an observed outcome $Y(\bs)$ and a treatment indicator $A(\bs)$,  $Y(\bs) = Y_0(\bs)\left\{1-A(\bs)\right\} + Y_1(\bs)A(\bs)$ for all $\bs\in\calD$.
    \label{sutva-std}
\end{assumption}
\begin{assumptionp}
    Ignorability: $A(\bs) \perp\!\!\!\perp \left\{{Y}_0(\bs), {Y}_1(\bs)\right\} \text{ given } \bX(\bs)$ for all $\bs\in\calD$.
    \label{ignor-std}
\end{assumptionp}
\begin{assumptionp}
    Positivity: $0<\mathbb{{P}}\{A(\bs) = 1 \mid \bX(\bs)\}<1$ almost surely for all $\bs\in\calD$.
    \label{posi_int-std}
\end{assumptionp}
These are the standard assumptions used in several causal inference frameworks \citep{hernan2010causal, imbens2015causal}. Assumption~\ref{sutva-std} states that ``there is no interference and there is only a single version of treatment" \citep{rubin1978bayesian}. Interference in our setting refers to a phenomenon in which the treatment or policy status at one location affects the potential outcomes of another. Therefore, this assumption implies that the observed outcomes correspond to the potential outcomes under the policy status of the location where they are measured, which is implicit in the potential outcome definition above that $Y_a(\bs)$ depends only on the treatment at $\bs$. Assumption~\ref{ignor-std} is also referred to as the no-unmeasured-confounders assumption \citep{rubin1978bayesian, rosenbaum1983central}, which is plausible if we have an ample set of covariates that are related to the outcome and policy assignment. However, this assumption is known to be untestable. Lastly, Assumption~\ref{posi_int-std} guarantees a sufficient covariate overlap between the two policy groups.}

    

\textcolor{black}{While Assumption \ref{sutva-std} is satisfied, Assumptions \ref{ignor-std} and \ref{posi_int-std} do not suffice in the presence of preferential sampling or spatial confounders.}  To model \textcolor{black}{possible} preferential sampling, define $\calS_a$ as the spatial point process for sampling locations for policy $a$, characterized by intensity $\lambda_a(\bs)$. The sampling processes  $\calS_a$ are potentially confounded by underlying spatial processes denoted $\calU_a = \left\{U_a(\bs):\bs\in\calD\right\}$ for $a\in\{0,1\}$ \textcolor{black}{that are shared components of the potential outcome processes \citep{diggle2010geostatistical}}. This results in \textcolor{black}{Assumption~\ref{ignor-std}} being violated, \textcolor{black}{meaning causal effects cannot be identified}. \textcolor{black}{Note that $U_a(\bs)$ are not potential outcomes of the latent processes, but rather unobserved spatial covariates that affect both the outcome and point process of each policy group (not affected by the policy status).} 
Similarly to $\calU_a$, define $\calV_a = \left\{V_a(\bs):\bs\in\calD\right\}$ as residual spatial processes to capture spatial trends within each point process not explained by $\calX$ and $\calU_a$. Let \textcolor{black}{$\calU=\calU_0 \cup \calU_1$, $\calV=\calV_0 \cup \calV_1$}, and $\mathcal{Y}_a = \left\{Y_a(\bs):\bs\in\calD\right\}$ for $a\in\{0,1\}$. 

In addition to Assumption \ref{sutva-std}, we require the following assumptions to identify the causal effects under preferential sampling or spatial confounding:
\begin{assumption}
    Latent ignorability: $(\calS_0, \calS_1) \perp\!\!\!\perp (\mathcal{Y}_0, \mathcal{Y}_1) \text{ given } \left\{\calX, \calU, \calV\right\}$.
    \label{ignor}
\end{assumption}
\begin{assumption}
    \textcolor{black}{Positivity}: $\lambda_a(\bs)>0$ almost surely for all $\bs\in\calD$ and $a\in\left\{0,1\right\}$.
    \label{posi_int}
\end{assumption}

\textcolor{black}{Assumption~\ref{ignor} implies the conditional independence of the potential outcomes from sampling mechanisms, including treatment allocation.} This relaxes \textcolor{black}{Assumption~\ref{ignor-std}} by allowing latent spatial processes to capture unmeasured confounders. \textcolor{black}{However, there is dependence between the sampling and potential outcomes processes marginally over the latent variables $\calU$ and $\calV$.} 
\textcolor{black}{For simplicity, we assume that $\calX$ is the set of confounders for modeling both the treatment allocation mechanism and sampling location processes. In general, in terms of mitigating bias, it is better to include variables that are unimportant than it is to miss important variables. 
} 
Like the standard Assumption \ref{ignor-std}, Assumption~\ref{ignor} is untestable in our case. 
Assumption~\ref{posi_int} declares that 
both treatments are possible for all $\bs\in\calD$, which enables us to estimate the local causal effects. \textcolor{black}{This is a modification of the standard positivity assumption given in Assumption~\ref{posi_int-std}.} This assumption can be checked using estimates of $\lambda_a(\bs)$.

We propose a hierarchical modeling framework for spatial causal inference in the presence of preferential sampling. The response model is specified in terms of potential outcomes \citep{rubin1974estimating}. 
Then the potential outcome model for each policy is
\begin{equation}
    Y_a(\bs) = \alpha_a + \bX(\bs)^{\top}\bbeta_a + U_a(\bs) + \epsilon_a(\bs),
    \label{po.model}
\end{equation}
where $\alpha_a$ is an intercept, $\bX(\bs)$ is a $p$-dimensional covariate vector, $\bbeta_a\in\mathbb{R}^p$ is a vector of regression coefficients, $\bU(\bs) = \left(U_0(\bs), U_1(\bs)\right)^{\top}$ is \textcolor{black}{modeled by} a mean-zero bivariate Gaussian process which might contain unmeasured spatial confounders, and $\epsilon_a(\bs)$ 
is a measurement error term distributed normally with mean zero and variance $\tau_a^2$, independent of $U_a(\bs)$. With this model specification, the local causal effect \eqref{local_causal_continuous} is defined as $$\Delta(\bs) = \alpha_1 - \alpha_0 + \bX(\bs)^{\top}\left(\bbeta_1 - \bbeta_0\right) + U_1(\bs) - U_0(\bs)$$
which is the conditional expectation of $Y_1(\bs)-Y_0(\bs)$ given $\left\{\calX, \calU, \calV\right\}$ \textcolor{black}{and averaging over local independent errors $\epsilon_a(\bs)$.}

To account for possible preferential sampling, we model the measurement locations using spatial point processes. 
A natural model for $\mathcal{S}_a$ is the doubly stochastic inhomogeneous Poisson point process \citep{pati2011bayesian} with log intensity
\begin{equation}
    \label{intensity.original}
    \log\left\{\lambda_a(\bs)\right\} = \eta_a^* + \bX(\bs)^{\top}\bdelta_a^* + V_a(\bs) + \phi_a \left\{\alpha_a + \bX(\bs)^{\top}\bbeta_a + U_a(\bs)\right\},
\end{equation}
where $\eta_a^*$ is an intercept, $\bdelta_a^*\in\mathbb{R}^p$ is a regression coefficient vector for each intensity model, and $\bV(\bs) = \left(V_0(\bs), V_1(\bs)\right)^{\top}$ is \textcolor{black}{modeled by} a mean-zero bivariate Gaussian process, which is independent of $\bU(\bs)$.
By incorporating $\alpha_a+\bX(\bs)^{\top}\bbeta_a + U_a(\bs)$ from (\ref{po.model}) in the intensity function, we allow for preferential sampling. That is, if $\phi_a>0$, the sites with larger mean response are more likely to be sampled. Under this formulation, the sampling locations under both policies follow log Gaussian Cox processes \citep{moller1998log, banerjee2003hierarchical}.

In this hierarchical model, $U_0(\bs)$ and $U_1(\bs)$ are shared latent processes between potential outcome models and point process models \citep{diggle2010geostatistical} to adjust for preferentially sampled locations, and $\phi_0$ and $\phi_1$ are coefficients to determine the degree of preferential sampling under policies 0 and 1, respectively. Under this model specification, the sampling locations are non-ignorable in computing the APE in \eqref{ate} due to the shared processes $U_0(\bs)$ and $U_1(\bs)$. 

We can reparameterize \eqref{intensity.original} by setting $\eta_a = \eta_a^* + \phi_a\alpha_a$ and $\bdelta_a = \bdelta_a^* + \phi_a\bbeta_a$ and specify the models for the point processes as
\begin{equation}
    \label{intensity}
    \log\left\{\lambda_a(\bs)\right\} = \eta_a + \bX(\bs)^{\top}\bdelta_a + V_a(\bs) + \phi_a U_a(\bs) + \psi_a(\bs).
\end{equation}
Given this specification, location densities ${\lambda_a(\bs)}/{\int_\calD \lambda_a(\bw)d\bw}$ are written as
\begin{equation}
    p_a\left(\bs|\mathcal{X}, \mathcal{U}_a, \mathcal{V}_a\right) = \frac{\exp\left\{\bX(\bs)^{\top}\bdelta_a + V_a(\bs) + \phi_a U_a(\bs)\right\}}{\int_{\calD} \exp\left\{\bX(\bw)^{\top}\bdelta_a + V_a(\bw) + \phi_a U_a(\bw)\right\}d\bw}
    \label{sampling_density}
\end{equation}
for $a\in\left\{0,1\right\}$.
By superposition of two Poisson processes \citep{durrett1999essentials}, given that a sample is taken at $\bs$, the probability of policy assignment mechanism conditioned on $\bX(\bs)$, $\bU(\bs)$, and $\bV(\bs)$ is
\begin{align}
    \mathbb{P}&\left\{A(\bs)=1 | \bX(\bs), \bU(\bs), \bV(\bs)\right\} = \frac{\lambda_1(\bs)}{\lambda_0(\bs) + \lambda_1(\bs)} = \text{expit}\left\{\mathbf{D}(\bs) \right\},
    \label{propscore}
\end{align}
where $\text{expit}(\cdot)$ is a sigmoid function and $$\bD(\bs) = \eta_1 - \eta_0 + \bX(\bs)^{\top}\left(\bdelta_1 - \bdelta_0\right) + \left\{V_1(\bs) + \phi_1 U_1(\bs)\right\} - \left\{V_0(\bs) + \phi_0 U_0(\bs)\right\}.$$ The probability defined in \eqref{propscore} is the propensity score \citep{rosenbaum1983central}. \textcolor{black}{From this expression, Assumption~\ref{ignor} implies that the propensity score given $\calX$, $\calU$, $\calV$, $\mathcal{Y}_0$ and $\mathcal{Y}_1$ depends only on  $\bX(\bs)$, $\bU(\bs)$ and $\bV(\bs)$, i.e., 
\begin{equation}
    \mathbb{P}\left\{A(\bs)=1|\mathcal{X}, \mathcal{U}, \mathcal{V}, \mathcal{Y}_0, \mathcal{Y}_1\right\} = \mathbb{P}\left\{A(\bs)=1|\bX(\bs), \bU(\bs), \bV(\bs)\right\}.
\end{equation}
Subsequently, $\left\{Y_0(\bs), Y_1(\bs)\right\}$ is conditionally independent of $A(\bs)$. Assumption~\ref{posi_int} ensures that 
\begin{equation*}
    \mathbb{P}\left\{A(\bs)=1|\bX(\bs), \bU(\bs), \bV(\bs)\right\}=\frac{\lambda_1(\bs)}{\lambda_0(\bs)+\lambda_1(\bs)}\in(0,1)
\end{equation*}
almost surely for all $\bs\in\calD$, guaranteeing that the propensity score remains strictly between 0 and 1.}

\begin{figure}[t]
    \centering
    \begin{tikzpicture}[
    scale = 2.0,
    roundnode/.style={circle, draw=black!60, thick, minimum size=7mm}
    ]
        \node[roundnode] (x) at (2.5,1) {$\calX$};
        \node[roundnode] (a) at (0,0) {$\calA$};
        \node[roundnode,dashed] (u) at (0,1) {$\calU$};
        \node[roundnode,dashed] (v) at (0,-1) {$\calV$};
        \node[roundnode] (y) at (2.5,0) {$\calY$};
        \node[roundnode] (s) at (2.5,-1) {$\calS$};
    
        \path[->] (x) edge (y);
        \path[->] (x) edge (a);
        \path[->] (a) edge (y);
        \path[->] (u) edge (y);
        \path[->] (u) edge (a);
        \path[->] (v) edge (a);
        \path[->] (v) edge (s);
        \path[->] (u) edge (s);
        \path[->] (x) edge[bend left] (s);
    \end{tikzpicture}
    
    \caption{\textcolor{black}{A directed acyclic graph (DAG) representing the causal and sampling structure of the proposed model. Here, $\calX=\left\{\bX(\bs):\bs\in \calD\right\}$ denotes observed covariates, $\calU=\left\{U_a(\bs):a\in{0,1}, \bs\in \calD\right\}$ latent spatial processes acting as unmeasured confounders for the outcome and sampling mechanisms, $\calV=\left\{V_a(\bs):a\in{0,1}, \bs\in \calD\right\}$ additional latent spatial effects specific to the sampling processes, $\calS = \calS_0\cup \calS_1$ the point pattern of sampled locations, $\calA=\left\{A(\bs):\bs\in \calD\right\}$ the policy (treatment) assignment, and $\calY=\left\{Y(\bs):\bs\in \calD\right\}$ the observed outcomes. Directed edges indicate conditional dependence implied by the hierarchical model, i.e., $\calU$ are confounders because the conditional distribution of the treatment and outcome processes depend on $\calU$. 
    Dashed nodes indicate unobserved processes. The graph reduces to a standard causal DAG under Assumptions (i) -- (iii) when $\calU$ is absent.}
    }
    \label{dag}
\end{figure}


\begin{figure}
    \centering
    \begin{subfigure}{.48\textwidth}
        \begin{tikzpicture}[
            scale = 0.8,
            node distance=1.2cm and 0.8cm, 
            >=Latex, 
            every node/.style={align=center}, 
            font=\footnotesize
        ]

        \node (popX) {Population of\\$\{Y_0(\bs) : \bs \in \calD\}$};
        \node[right=of popX] (popY) {Population of\\$\{Y_1(\bs) : \bs \in \calD\}$};

        \node[below=of popX] (Z0) {$\calS_0$};
        \node[below=of popY] (Z1) {$\calS_1$};



        \node[below=of Z0] (obsX) {$\{Y_0(\bs) : \bs \in \calS_0\}$};
        \node[below=of Z1] (obsY) {$\{Y_1(\bs) : \bs \in \calS_1\}$};

        \draw[->] (popX) -- (Z0);
        \draw[->] (popY) -- (Z1);

        \node[right=0.1cm] at ($(popX)!0.5!(Z0)$) {Sample\\with $\lambda_0(\bs)$};
        \node[right=0.1cm] at ($(popY)!0.5!(Z1)$) {Sample\\with $\lambda_1(\bs)$};

        \node[right=0.1cm] at ($(Z0)!0.4!(obsX)$) {Observe at\\$\calS_0$};
        \node[right=0.1cm] at ($(Z1)!0.4!(obsY)$) {Observe at\\$\calS_1$};


        \draw[->] (Z0) -- (obsX);
        \draw[->] (Z1) -- (obsY);


        \coordinate (centerBelow) at ($(obsX.south west)!0.5!(obsY.south east)$);
        
        \node[below=0.8cm of centerBelow] (finalSet) {$\text{Observation:}$\\$\{\bX(\bs), A(\bs), Y(\bs) : \bs \in \calS_0 \cup \calS_1\}$};
        \draw[->] (obsX) -- (finalSet);
        \draw[->] (obsY) -- (finalSet);
        \node[left=0.1cm] at ($(finalSet)!0.5!(obsX)$) {$A(\bs)=0$};
        \node[right=0.1cm] at ($(finalSet)!0.5!(obsY)$) {$A(\bs)=1$};
    \end{tikzpicture}
    \caption{\textcolor{black}{Path 1: Policy-specific point processes for sampling locations.}}
    \label{schematic1}
    \end{subfigure}
    \hfill
    \begin{subfigure}{.48\textwidth}
        \centering
        \begin{tikzpicture}[
            scale = 0.8,
            node distance=1.9cm and 1.2cm, 
            >=Latex, 
            every node/.style={align=center}, 
            font=\footnotesize
        ]

        \node (popX) {Population of\\$\{Y_a(\bs) : a\in\{0,1\}, \bs \in \calD\}$};
        \node[below=of popX] (obsX) {$\left\{Y(\bs) = Y_1(\bs)A(\bs) + Y_0(\bs)\left\{1-A(\bs)\right\}:\bs\in\calD\right\}$};
        \node[right=0.1cm] at ($(popX)!0.5!(obsX)$) {$A(\bs)=1$ with\\$\frac{\lambda_1(\bs)}{\lambda_0(\bs)+\lambda_1(\bs)}$};
        \node[below=of obsX] (finalSet) {
        $\text{Observation:}$\\$\{\bX(\bs), A(\bs), Y(\bs) : \bs \in \calS\}$};
        \node[right=0.1cm] at ($(obsX)!0.4!(finalSet)$) {Sample $\calS$\\with $\lambda_0(\bs) + \lambda_1(\bs)$};
        \draw[->] (obsX) -- (finalSet);
        \draw[->] (popX) -- (obsX);
    \end{tikzpicture}
    \caption{\textcolor{black}{Path 2: A spatial exposure process with identical and independent sampling of locations }}
    \label{schematic2}
    \end{subfigure}
    \caption{\textcolor{black}{Two equivalent representations of the data-generating mechanism. (a) Sampling-first formulation: locations are generated from two policy-specific point processes with intensities $\lambda_0(\bs)$ and $\lambda_1(\bs)$, yielding samples $\calS_0$ and $\calS_1$. (b) Exposure-first formulation: the policy assignment is generated with probability $\lambda_1(\bs)/{\left\{\lambda_0(\bs)+\lambda_1(\bs)\right\}}$, and then locations are first sampled from the superposed process with intensity $\lambda_0(\bs)+\lambda_1(\bs)$. Both formulations induce the same joint distribution for $(\calS, \calA, \calY)$ conditional on the latent spatial processes.}}
\end{figure}

\textcolor{black}{A directed acyclic graph (DAG) representing the causal and sampling structure of the proposed model is found in Figure~\ref{dag}. The proposed model is formulated in terms of policy-specific point processes for sampling locations, as illustrated in Figure~\ref{schematic1}. Under this formulation, locations in the controlled and treated groups are generated separately according to intensities $\lambda_0(\bs)$ and $\lambda_1(\bs)$, respectively, and the observed dataset consists of $\left\{\bX(\bs),A(\bs),Y(\bs)\right\}$ at the union $\calS=\calS_0\cup \calS_1$. The DAG highlights the role of $\calU$ as the unmeasured confounders that appear in the conditional distribution of the treatment and outcome processes.}

\textcolor{black}{This representation is mathematically equivalent to an alternative sequential formulation shown in Figure~\ref{schematic2}. In this view, the treatment assignment is generated according to the propensity score in \eqref{propscore}, and then locations are sampled from the superposed point process with intensity $\lambda_0(\bs)+\lambda_1(\bs)$. Conditional on the latent spatial processes $\calU$ and $\calV$, sampling and treatment assignment are independent across locations, but they are marginally dependent through these shared processes. Similar dual representations arise in survey sampling models with nonresponse and informative sampling \citep{deville1994variance, shao1999variance}.}


Turning to the modeling of the bivariate spatial random effects, we employ a linear model of coregionalization (LMC) \citep{banerjee2003hierarchical}. 
Define $\Tilde{U}_{a}(\bs)$ for $a\in\left\{0,1\right\}$ as independent Gaussian processes with Matérn covariance function $\sigma_{u,a}^2\calR\left(h;\rho_u,\kappa_u\right)$ \citep{williams2006gaussian}, where $h=\|\bs-\bs'\|$ for $\bs,\bs'\in\calD$, $\rho_u > 0$ is a spatial dependence parameter, $\kappa_u > 0$ is a smoothness parameter, and
\begin{equation}
    \calR\left(h;\rho_u,\kappa_u\right) = \frac{1}{2^{\kappa_u-1}\Gamma(\kappa_u)}\left(\frac{h}{\rho_u}\right)^{\kappa_u}\mathcal{K}_{\kappa_u}\left(\frac{h}{\rho_u}\right).
    \label{matern}
\end{equation}
Here, $\mathcal{K}_{\kappa_u}$ denotes the modified Bessel function of the second kind. A special case of $\calR(h;\rho_u, \kappa_u)$ is an exponential correlation $\calR(h;\rho_u) = \exp\left(-h/\rho_u\right)$ with $\kappa_u=1/2$.
Then, using the LMC, we have
\begin{align*}
    U_{0}(\bs) = \Tilde{U}_{0}(\bs) \hspace{5pt}\text{ and }\hspace{5pt} U_{1}(\bs) = \Tilde{U}_{1}(\bs) + \gamma_u \Tilde{U}_{0}(\bs)
\end{align*}
for all $\bs\in\calD$. The bivariate spatial processes $V_0(\bs)$ and $V_1(\bs)$ are modeled analogously.

To understand the consequences of preferential sampling, we evaluate the expectation of $Y_1(\bs) - Y_0(\bs)$ averaging over both the spatial location and potential outcomes, conditionally on $\calX$. Supplemental material S1 shows that $\mbox{E}\{Y_1(\bs)-Y_0(\bs)\}$ is approximated as
\begin{align}
    (\alpha_1-\alpha_0) + \left({\bar\bX}_1^{\top}\bbeta_1 - {\bar\bX}_0^{\top}\bbeta_0\right) + \left[\phi_1\text{Var}\left\{U_1(\bs)\right\} - \phi_0\text{Var}\left\{U_0(\bs)\right\}\right]\label{e:bias}, 
\end{align}
where ${\bar\bX}_a = \int_\calD \bX(\bw)\exp\{\bX(\bw)^{\top}\bdelta_a\}d\bw/\int_\calD \exp\{\bX(\bw)^{\top}\bdelta_a\}d\bw$, and the variances of the stationary processes are $\text{Var}\left\{U_1(\bs)\right\} = \sigma_{u,1}^2 + \gamma_u^2\sigma_{u,0}^2$ and $\text{Var}\left\{U_0(\bs)\right\} = \sigma_{u,0}^2$. The second term shows how preferential sampling affects the covariate distribution in the two treatment groups.  The third term quantifies the effects of preferential sampling that cannot be explained by covariates. When $\text{Var}\left\{U_0(\bs)\right\} = \text{Var}\left\{U_1(\bs)\right\} = 0$ or $\phi_0=\phi_1=0$, this term is zero, meaning the underlying spatial processes are deterministic or the sampling intensities do not depend on the response.  This expression also suggests that the effects of preferential sampling are the most prominent when the sampling protocols differ by treatment group, e.g., if $\text{Var}\left(U_0\right) = \text{Var}\left(U_1\right)$ then the term is proportional to $\phi_1-\phi_0$, and so the term is nonzero only if the strength of preferential sampling differs by treatment group.


\section{Computational and theoretical properties of the proposed model}
We propose a fully Bayesian analysis using Markov Chain Monte Carlo (MCMC) to estimate $\triangle$ and its uncertainty. Computation is complicated by the stochastic integration in \eqref{sampling_density}. In Section~\ref{car_detail}, we describe a discretized model to approximate this integration, and in Section~\ref{mcmc}, we provide MCMC details to fit this discretized model. \textcolor{black}{In Section~\ref{sec:theory}, we provide the results of the parameter identification and the posterior consistency of the proposed model.}

\subsection{Grid approximation}\label{car_detail}
The proposed model consists of stochastic integrals making computation challenging. As such, numerical integration using grid cells is employed \citep{diggle2010geostatistical, pati2011bayesian, gelfand2018bayesian}.  
We partition the domain into $G$ equally-sized grid cells, i.e., $\calD = \cup_{g=1}^G\calD_g$ with $|\calD_g| = \frac{|\calD|}{G}$ for all $g=1,\cdots,G$, and approximate the integral as
\begin{equation}
    \int_{\calD} \exp\left\{\bX(\bs)^{\top}\bdelta_a + V_a(\bs) + \phi_aU_a(\bs)\right\}d\bs \approx \frac{|\calD|}{G}\sum_{g=1}^G\exp\left(\bX_g^{\top}\bdelta_a + V_{a,g} + \phi_aU_{a,g}\right)
    \label{approx}
\end{equation}
for $a\in\left\{0,1\right\}$. In \eqref{approx}, $\bX_g$, $U_{a,g}$ and $V_{a,g}$ are grid-level values of $\bX(\bs)$, $U_a(\bs)$, and $V_a(\bs)$ evaluated at the centroid of the $g^{\text{th}}$ grid cell $\bc_g$, i.e., $\bX_g = \bX(\bc_g)$ and similarly for $U_{a,g}$ and $V_{a,g}$. This implies that the spatial random effects and the covariates representing each grid cell are assumed to be constant in each grid cell.

By using the LMC and the property of Gaussian processes that any finite collection of them follows a multivariate normal distribution (MVN), we have
\begin{align*}
    U_{0,g} = \Tilde{U}_{0,g} \hspace{5pt}\text{ and }\hspace{5pt} U_{1,g} = \Tilde{U}_{1,g} + \gamma_u \Tilde{U}_{0,g}
\end{align*}
for $g=1,\cdots,G$ where $\Tilde{\mathbf{U}}_a = \left(\Tilde{U}_{a,1}, \cdots, \Tilde{U}_{a,G}\right)^{\top}$ is distributed as
\begin{equation}
    \Tilde{\bU}_a \sim \text{MVN}\left\{\mathbf{0}_G, \sigma_{u,a}^2\bR(\rho_u, \kappa_u)\right\}.
    \label{caru}
\end{equation}
Here, $\mathbf{0}_G$ is a zero vector of length $G$ and $\bR(\rho_u, \kappa_u)$ is a correlation matrix with its diagonal entries being $1$ and off-diagonal elements corresponding to $\calR(\|\bc_g-\bc_{g'}\|;\rho_u,\kappa_u)$ for $g,g'\in\left\{1,\cdots,G\right\}$. Using a similar argument, $V_{0,g}$ and $V_{1,g}$ can be modeled by $V_{0,g} = \Tilde{V}_{0,g}$ and $V_{1,g} = \Tilde{V}_{1,g} + \gamma_v \Tilde{V}_{0,g}$ for $g=1,\cdots, G$, where $\Tilde{\bV}_{a}\sim\text{MVN}\left\{\mathbf{0}_G, \sigma_{v,a}^2\bR(\rho_v, \kappa_v)\right\}$ for $a\in\left\{0,1\right\}$.

Given the spatial random effects, the number of observations within each grid cell follows a Poisson distribution. The hierarchical model is written as
\begin{align}
&Y_{0}(\bs) = \alpha_0 + \bX(\bs)^{\top}\bbeta_0 + \Tilde{U}_{0,g(\bs)} +\epsilon_{0}(\bs) \label{hierar-car1}  \\
    &Y_{1}(\bs) = \alpha_1 + \bX(\bs)^{\top}\bbeta_1 + \Tilde{U}_{1,g(\bs)} + \gamma_u\Tilde{U}_{0,g(\bs)}  +\epsilon_{1}(\bs) \label{hierar-car2} \\
    &N_{0,g} \sim \text{Poisson}\left(|\calD_g|\cdot\lambda_{0,g} \right) \label{hierar-car3} \\
    &N_{1,g} \sim \text{Poisson}\left(|\calD_g|\cdot\lambda_{1,g}\right) \label{hierar-car4} \\
    &\log\left\{\lambda_{0,g}\right\} = \eta_0 + \bX_g^{\top}\bdelta_0 + \Tilde{V}_{0,g} + \phi_0 \Tilde{U}_{0,g} + \psi_{0,g} \label{hierar-car5} \\
    &\log\left\{\lambda_{1,g}\right\} = \eta_1 + \bX_g^{\top}\bdelta_1 + \Tilde{V}_{1,g} + \gamma_v\Tilde{V}_{0,g} + \phi_1\left(\Tilde{U}_{1,g} + \gamma_u\Tilde{U}_{0,g}\right) + \psi_{1,g}\label{hierar-car6}
\end{align}
for $g=1,\cdots,G$, where $g(\bs)$ is the index of the grid where $\bs$ lies, i.e., $g(\bs) = k$ for some $k\in\left\{1,\cdots,G\right\}$ if $\bs\in\calD_{k}$. Here, $N_{a,g}$ is the number of observations in $g$ which received treatment $a\in\left\{0,1\right\}$, and $\psi_{a,g}$ is an additional random effect component for explaining nonspatial heterogeneity, which is independently distributed as $\text{Normal}(0, \tau_{\psi_a}^2)$. Note that $\sum_{g=1}^G N_{a,g} = n_a$ is the total number of samples taken at locations under policy $a$, and $n_0 + n_1 = n$ is the sample size. 

The APE, $\triangle$, over the domain is approximated by
\begin{equation}
    \triangle \approx G^{-1}\sum_{g=1}^{G} \triangle_g,
    \label{ate_approx}
\end{equation}
where $\triangle_g = \alpha_1 - \alpha_0 + \bX_g^{\top}\left(\bbeta_1 - \bbeta_0\right) + U_{1,g} - U_{0,g}$ is a local causal effect at grid cell $\calD_g$. If the grid cell sizes vary, \eqref{ate_approx} can be modified by applying weights. The propensity score at $\calD_g$ is approximated as
\begin{equation}
    \frac{\lambda_{1,g}}{\lambda_{0,g} + \lambda_{1,g}} = \text{expit}\left\{\eta_1 - \eta_0 + \bX_g^{\top}\left(\bdelta_1 - \bdelta_0\right) + \left(V_{1,g} + \phi_1 U_{1,g}\right) - \left(V_{0,g} + \phi_1 U_{0,g}\right)\right\}.
    \label{grid.prop}
\end{equation}

\subsection{MCMC details}\label{mcmc}
Model inference is obtained in a Bayesian framework.
We assign proper conjugate Gaussian priors for $\alpha_a$, $\bbeta_a$, $\eta_a$, and $\bdelta_a$ for $a\in\left\{0,1\right\}$. Normal conjugate priors are assigned to the parameters controlling the degree of preferential sampling, \textcolor{black}{$\phi_0$ and $\phi_1$}, and to the parameters of the LMC, \textcolor{black}{$\gamma_u$ and $\gamma_v$}. Inverse-gamma priors are assigned to the variance components $\tau_a^2$, $\sigma_{u,a}^2$, $\sigma_{v,a}^2$, and $\tau_{\psi_a}^2$ for $a\in\left\{0,1\right\}$. We assign proper priors to the spatial dependence parameters and the smoothness parameters $\rho_u$, $\rho_v$, $\kappa_u$ and $\kappa_v$. The spatial random effects $\Tilde{\bU}_a$ and $\Tilde{\bV}_a$ are modeled using the Gaussian processes as described in Section~\ref{car_detail}. 

Bayesian inference of the parameters and hyperparameters for the models~\eqref{hierar-car1}-\eqref{hierar-car6} is obtained using MCMC. 
The model parameters and hyperparameters are sampled using a mixture of Gibbs sampling, Metropolis random walk, and Hamiltonian Monte Carlo (HMC) sampling \citep{neal2011mcmc}. Missing potential outcomes are imputed using the predictive distributions \eqref{hierar-car1} and \eqref{hierar-car2} at every step of MCMC. 
The log intensities are updated by HMC, while the covariance hyperparameters $\rho_u$, $\rho_v$, $\kappa_u$ and $\kappa_v$ are updated via Metropolis random walk. All the remaining parameters are updated by Gibbs samplers. The details of the MCMC algorithm can be found in Supplemental Material.

\subsection{Theoretical properties of the proposed model}\label{sec:theory}

In this section, we establish the theoretical properties of the proposed method. We first declare additional notation and assumptions. We define $\btheta = \left(\btheta_\calM^{\top}, \btheta_\calC^{\top}\right)^{\top}$ where
\begin{equation}
    \btheta_\calM = \left\{\alpha_0, \bbeta_0, \bdelta_0, \eta_0, \phi_0, \tau_0, \alpha_1, \bbeta_1, \eta_1, \bdelta_1, \phi_1, \tau_1, \gamma_u, \gamma_v\right\}
    \label{parameter}
\end{equation}
is the vector of parameters of interest and
\begin{equation}
    \btheta_\calC = \left\{\rho_u, \rho_v, \kappa_u, \kappa_v, \sigma_{u,0}^2, \sigma_{u,1}^2, \sigma_{v,0}^2, \sigma_{v,1}^2\right\}
    \label{cov_parameter}
\end{equation}
represents the vector of hyperparameters. Additionally, we define 
\begin{equation}
    {\calW} = \left\{\left(\Tilde{U}_a(\bs), \Tilde{V}_a(\bs)\right):\bs\in\calD, a\in\left\{0,1\right\}\right\}.
\end{equation} 

The theorems require the following assumptions.
\begin{assumption}
\label{spatial_process}
    The prior density for $\btheta_\calM$ is positive for all possible values of $\btheta_\calM$. Further, the prior for $\kappa_u$ satisfies $\mathbb{P}\left\{\kappa_u \geq 1/2\right\} > 0$ (similarly for $\kappa_v$), while proper priors are assigned to $\rho_u$, $\rho_v$, $\sigma_{u,a}^2$ and $\sigma_{v,a}^2$ for $a\in\left\{0,1\right\}$.
\end{assumption}
\begin{assumption}\label{linear_indep}
    The matrices $\bX_\calS$ with $i^{\text{th}}$ row corresponding to $\bX(\bs_i)$ and $\bX_\calG$ with $k^{\text{th}}$ row corresponding to $\bX(\bc_k)$ are full rank.
\end{assumption}
\begin{assumption}
    $\bX(\bs)$ is uniformly bounded, i.e., there exists $C>0$ such that $\|\bX(\bs)\| \leq C$ for all $\bs\in\calD$.
    \label{unif-bdd}
\end{assumption}

Finally, the following theorems hold.
\begin{theorem}
    Under models \eqref{hierar-car1} -- \eqref{hierar-car6} and Assumptions~\ref{sutva-std} -- \ref{posi_int} and \ref{linear_indep}, the vector of parameters $\btheta$ is identifiable.
    \label{identifiability}
\end{theorem}

\begin{theorem}
    Under the models \eqref{po.model} and \eqref{intensity}, and Assumptions~\ref{sutva-std} -- \ref{spatial_process}, and \ref{unif-bdd}, the posterior distribution of $\bomega = (\btheta_\calM, \calW)$
    is weakly consistent.
    \label{weak-consistency}
\end{theorem}
\noindent
Theorem~\ref{identifiability} guarantees that the model parameters are identifiable without any information from the prior distributions for $\btheta$. In particular, we can identify the correlation between potential outcomes at the same location by leveraging spatial dependence, which is not possible for non-spatial models. Theorem~\ref{weak-consistency} states that within a class of prior distributions for $\btheta_\calM$ and $\calW$, the posterior converges to the true distribution in a weak sense, which implies that $\triangle(\bs)$ and $\triangle$ are estimable. Refer to \cite{ghosh2003nonpara} or \cite{ghosal2017fundamentals} for a formal definition of weak consistency in detail. Proofs are given in Supplemental Material.

\section{Simulation study}\label{sec:sim}

In this section, we conduct simulation studies to illustrate the bias caused by ignoring preferential sampling and the ability of the proposed method to mitigate this bias. We compare the performance of the proposed hierarchical model (``full") with the model that only considers the response layers \eqref{hierar-car1} and \eqref{hierar-car2} and thus ignores preferential sampling (``naive"). We also \textcolor{black}{consider} the model proposed by \cite{pati2011bayesian} (``shared"), which is defined
\begin{align*}
    &Y(\bs) = {\alpha}_{0,\text{P}} + \alpha_{1,\text{P}}\cdot A(\bs) + \bX(\bs)^{\top}\bbeta_{0,\text{P}} + A(\bs)\cdot\bX(\bs)^{\top}\bbeta_{1,\text{P}} + {U}_{g(\bs)} +\epsilon(\bs)   \\
    &N_{g} \sim \text{Poisson}\left(|\calD_g|\cdot\lambda_{g}\right)  \\
    &\log\left\{\lambda_{g}\right\} = \eta + \bX_g^{\top}\bdelta + {V}_{g} + \varphi{U}_{g} + \psi_{g}.
\end{align*}
This model assumes shared spatial processes, $U(\bs)$ and $V(\bs)$, between the two policy groups for both the response and the point process. Under the shared model with Assumption~\ref{sutva-std}--\ref{posi_int}, the APE is defined as $$\alpha_{1,P} + G^{-1}\sum_{g=1}^G \bX_g^{\top}\bbeta_{1,P}.$$ \textcolor{black}{Finally, we further consider the propensity score adjustment to the outcome regression model as a competitor. This two-stage procedure first estimates the propensity scores using either Bayesian additive regression trees \citep{chipman2010bart} (BART) or spatial generalized linear models. Then, B-spline basis functions are constructed using the estimated propensity scores with degrees of freedom $5$. These B-splines are included as covariates in the outcome regression models \eqref{hierar-car1} and \eqref{hierar-car2}. See \cite{hill2011bayesian}, \cite{alaa2018bayesian}, \cite{ray2020semiparametric}, or \cite{bae2024bayesian} for more information regarding Bayesian causal inference with nonparametric or semiparametric frameworks.}

\subsection{Data generation process}\label{s:sim:dgp}
The domain of interest is $\calD = [0,1]^2$, which we discretize into $20\times20$ square grid cells. The data generation is based on the model defined by \eqref{hierar-car1} -- \eqref{hierar-car6}. We generate $p=2$ independent covariates from Gaussian processes with exponential covariance kernel whose variance is $0.5$ and spatial dependence parameter is $0.05$, to construct $\bX_g$ for $g=1,\cdots,G=20^2$.

 We consider \textcolor{black}{ten} different scenarios to compare the \textcolor{black}{five} models. The first six scenarios differ based on (i) degree of preferential sampling, (ii)  spatial dependence, and (iii) the expected number of observations. The last two scenarios consider nonstationarity and non-Gaussian data. Specific details for each scenario are given in Table~\ref{t:sim_result_pate}. In each scenario, we set $\alpha_0 = 2$, $\alpha_1 = 4$, $\bbeta_0 = (1, 1)^{\top}$, $\bbeta_1 = (-1, -1)^{\top}$, $\bdelta_0^* = (1, 1)^{\top}$ and $\bdelta_1^* = (-1, -1)^{\top}$. Spatial random fields are generated by Gaussian processes with exponential covariance functions so that $\Tilde{\bU}_a$ and $\Tilde{\bV}_a$ are distributed as $\text{MVN}\left\{\mathbf{0}_G, \sigma^2\bR(\rho) \right\}$ where $(g,g')$ entry of $\bR(\rho)$ is $\exp\left(-\|\bc_g-\bc_{g'}\|/\rho\right)$, $\sigma^2 = 1$ and $\rho \in \left\{0.1, 0.2\right\}$ for $g,g'\in\left\{1,\cdots,20^2\right\}$. The LMC parameters are set to $\gamma_u = -0.5$ and $\gamma_v = 0.5$. The degrees of preferential sampling are set to be $\phi_0 = \phi$ and $\phi_1 = 1.5\phi$, where $\phi\in\left\{0, 1/3, 2/3, 1\right\}$, with $\phi = 0$ signifying nonpreferential sampling and $\phi = 1$ signifying strong preferential sampling. Recall from \eqref{intensity.original} that $\bdelta_a = \bdelta_a^* + \phi_a\bbeta_a$ varies with $\phi_a$.  The variances are all set to $\tau_0^2 = \tau_1^2 = \tau_{\psi_0}^2 = \tau_{\psi_1}^2 = 0.1$. The number of observations in each grid cell, $N_{0,g}$ and $N_{1,g}$, are generated from Poisson distributions where the mean intensities $\lambda^*=G^{-1}\sum_{g=1}^G \lambda_{a,g}$ are set to $\lambda^* \in \left\{5,10\right\}$. This is achieved by controlling the values of $\eta_0$ and $\eta_1$ for each simulation set. These values result in $4,000$ observations, on average, when $\lambda^*=5$, and $8,000$ observations, on average, when $\lambda^*=10$. The realizations of the Poisson process on each grid cell represent the number of observations located in the grid cell. We set the covariates of each observation to have the same values as those of the grid cell in which the site is located. We present six scenarios by varying these parameters as specified in Table~\ref{t:sim_result_pate}.
 
 To examine robustness to model misspecification, Scenarios 7 and 8 generate spatial processes that are nonstationary and non-Gaussian, respectively. The grid cell centroids with the $y-$coordinates being squared are used to generate nonstationary $\bV_0$ and $\bV_1$ (scenario 7).
 To incorporate non-Gaussianity (scenario 8), the log intensities of the following forms
$$\log\left\{\lambda_{0,g}\right\}=\eta_0 + \bX_g^{\top}\bdelta_0 + {V}_{0,g} + \phi_0\left\{U_{0,g}\mathbb{I}\left(U_{0,g}>0\right) - \sqrt{\frac{\sigma^2}{2\pi}}\right\} + \psi_{0,g} $$
and
$$\log\left\{\lambda_{1,g}\right\}=\eta_1 + \bX_g^{\top}\bdelta_1 + {V}_{1,g} + \phi_1\left\{U_{1,g}\mathbb{I}\left(U_{1,g}>0\right) - \sqrt{\frac{(1+\gamma_u^2)\sigma^2}{2\pi}}\right\} + \psi_{1,g}$$
are considered to introduce the effect of $\bU_0$ and $\bU_1$ on the point processes after non-Gaussian transformations. \textcolor{black}{The last two scenarios are included to examine performance of the models under different values of $\gamma_u$, and, subsequently, different levels of correlation between $Y_0(\bs)$ and $Y_1(\bs)$ at each location $\bs\in\calD$. In Scenario 9, $\gamma_u = 0$ results in independence between $Y_0(\bs)$ and $Y_1(\bs)$, while $\gamma_u = 0.5$ in Scenario 10 results in positive correlation between the two potential outcomes.}

\textcolor{black}{For the full and naive models,} we assign the proper priors $\alpha_a \sim \text{Normal}\left(0, 10^2\right)$, $\bbeta_a \sim \text{MVN}\left(\boldsymbol{0}_2, 10^2\mathbb{I}_2\right)$, $\eta_a \sim \text{Normal}\left(0, 10^2\right)$, and $\bdelta_a \sim \text{MVN}\left(\boldsymbol{0}_2, 10^2\mathbb{I}_2\right)$ for $a\in\left\{0,1\right\}$. The LMC parameters,$\gamma_u$ and $\gamma_v$, and the parameters governing preferential sampling, $\phi_0$ and $\phi_1$, are assigned independent $\text{Normal}\left(0, 10^2\right)$. The variance components $\tau_a^2$, $\sigma_{u,a}^2$, $\sigma_{v,a}^2$, and $\tau_{\psi_a}^2$ are assigned with $\text{Inverse-gamma}(0.1, 0.1)$ priors for $a\in\left\{0,1\right\}$. We assign $\text{Uniform}\left(0, 0.5\right)$ priors to the spatial dependence parameters $\rho$. \textcolor{black}{Similar priors are assigned to the counterparts of the shared models.} For each data set and scenario, MCMC is run for $120,000$ iterations with the first $50,000$ samples discarded as burn-in.

\subsection{Results}\label{s:sim:results}
 
We generate $200$ data sets for each scenario. The causal estimand of interest is $\triangle$ given in \eqref{ate_approx}. Table~\ref{t:sim_result_pate} reports bias, mean squared error (MSE) and 95\% coverage probability (CP) for $\triangle$ \textcolor{black}{under the naive and shared models, the outcome regression models with propensity score adjustments, and the full model.} Throughout all scenarios, the shared model shows the poorest performance compared to the full and naive models. We speculate that this is due to the single use of the latent spatial process in the shared model, which makes the model lack flexibility. Moreover, the shared model assumes the identical degree of covariate effects for both the treatment and control groups in the point processes. Expression~\eqref{e:bias} explains the extreme bias of the estimated APEs in this model, even when there is no preferential sampling, as shown in Scenario 1. Therefore, we will focus on the comparison between the full and naive models.

In comparing Scenarios 1 through 4, the discrepancy in model performance between the full and naive models becomes larger as the degree of preferential sampling increases (from $\phi=0$ to $\phi=1$). Specifically, in Scenario 1 without preferential sampling, both models show similar performance in terms of bias, MSE, and CP. With moderate preferential sampling in Scenario 2, the naive model exhibits poorer performance in terms of bias, MSE, and CP, while the full model shows almost similar performance compared to Scenario 1. In Scenarios 3 and 4 with strong and extreme preferential sampling, respectively, the full model outperforms the naive model in terms of bias, MSE, and CP. Specifically, the CPs of the naive models drop significantly, while both the biases and MSEs increase. In contrast, the values of the full model remain similar to those observed in Scenarios 1 through 3. This demonstrates the impact on the estimation of the causal effect when preferential sampling is not considered. \textcolor{black}{This tendency towards poor model performance with stronger degrees of preferential sampling prevails in the propensity score adjustment approaches. The models using adjustments from BART or spatial generalized linear models have similar performance to the naive model, indicating the propensity score adjustment is not effective in removing bias due to preferential sampling.}

We see in Scenario 5 that greater spatial dependence tends to reduce MSE when compared to Scenario 3 for both models. The full model still shows better performance than the naive model in terms of bias, MSE, and CP. In Scenario 6 where the total sample size is twice as large on average, the bias and MSE decreased for both the full and naive models compared to Scenario 3. The full model continues to outperform the naive model across all three measures. \textcolor{black}{Similar to the first four scenarios, the propensity score adjustment models have equivalent performance as the naive model for scenarios 5 and 6.}

In Scenario 7, the full model outperforms the naive model \textcolor{black}{and the propensity score adjustment models} in terms of bias, MSE, and CP, while the performance of the full model is similar to that in Scenario 3. In Scenario 8, the full model is shown to have similar performance to that under Scenario 3, exhibiting the robustness to the mis-specification of the point process models. However, the naive model \textcolor{black}{and the propensity score adjustment models} show similar performance compared to scenario 3 in terms of MSE and CP. Importantly, Scenarios 7 and 8 illustrate that the proposed model is robust to some forms of model misspecification. \textcolor{black}{In Scenario 9, the bias and MSE are lower than those of Scenario 3 or 10 for all models, and we observe similar improvements in the CP for this scenario. 
However, the full model still outperforms all other models in terms of the three criteria. The similar trend can be seen in Scenario 10, exhibiting that the full model is effective in reducing the preferential sampling bias under different values of $\gamma_u$.}

\begin{table}[t]
    \scriptsize
    {\renewcommand{\arraystretch}{1.11}%
    \begin{tabular}{c|cccccc|ccccc}
                \multicolumn{11}{c}{} \\
				
				&$\phi$ & $\rho$ & $\lambda^*$ & $\gamma_u$ & G & S & Naive & Shared & PSA-B & PSA-G & Full \\
				\hline
				1 &0.0 & 0.1 & 5 & -0.5 & \checkmark & \checkmark & -0.2 (0.3) & 0.6 (1.5) & -0.3 (0.3) & -0.1 (0.3) & -0.5 (0.3) \\
				2 &1/3 & 0.1 & 5 & -0.5 & \checkmark & \checkmark & 6.8 (0.5) & 21.5 (1.7) &  6.9 (0.5) & 6.9 (0.5) & -0.7 (0.6) \\
				3 &2/3 & 0.1 & 5 & -0.5 & \checkmark & \checkmark & 20.6 (0.8) & 47.3 (2.4) & 21.4 (0.9) & 20.4 (0.8) & -1.7 (0.8) \\
				4 &1.0 & 0.1 & 5 & -0.5 & \checkmark & \checkmark & 40.3 (1.5) & 74.2 (3.4)& 42.5 (1.9) & 38.7 (1.6) & -1.8 (1.5) \\
				5 &2/3 & 0.2 & 5 & -0.5 & \checkmark & \checkmark & 13.6 (0.7) & 40.7 (2.7) & 14.1 (0.8) & 13.7 (0.7) & -0.3 (0.6) \\
				6 &2/3 & 0.1 & 10 & -0.5 & \checkmark & \checkmark & 17.5 (0.7) & 47.0 (2.1) & 18.0 (0.7) & 17.4 (0.7) & -1.3 (0.6) \\
				7 &2/3 & 0.1 & 5 & -0.5 & \checkmark &  & 22.1 (0.9) & 50.3 (2.5) & 22.5 (1.0) & 21.6 (0.9) & 0.4 (0.9) \\
				8 &2/3 & 0.1 & 5 & -0.5 &  & \checkmark & 10.4 (0.8) & 35.7 (3.2) & 11.0 (0.9) & 10.6 (0.9) & -4.3 (0.8) \\
                9 &2/3 & 0.1 & 5 & 0.0 & \checkmark & \checkmark & 15.7 (0.8) & 25.6 (1.9) & 15.4 (0.9) & 15.5 (0.9) & -1.0 (0.8) \\
                10 &2/3 & 0.1 & 5 & 0.5 & \checkmark & \checkmark & 21.2 (0.8) & 29.2 (1.7) & 20.8 (0.9) & 21.2 (0.9) & -1.6 (0.9) \\
                \end{tabular}}
    \vspace{0.1cm}\\
    \text{(a) Bias}\\
    {\renewcommand{\arraystretch}{1.11}%
    \begin{tabular}{c|cccccc|ccccc}
				\multicolumn{11}{c}{} \\
				
				&$\phi$ & $\rho$ & $\lambda^*$ & $\gamma_u$ & G & S & Naive & Shared & PSA-B & PSA-G & Full\\
				\hline
				1 &0.0 & 0.1 & 5 & -0.5 & \checkmark & \checkmark & 0.2 (0.0) & 4.4 (0.4) & 0.2 (0.0) & 0.2 (0.0) & 0.2 (0.0) \\
				2 &1/3 & 0.1 & 5 & -0.5 & \checkmark & \checkmark & 1.0 (0.1) & 10.3 (0.9) & 1.0 (0.1) & 1.0 (0.1) & 0.6 (0.1) \\
				3 &2/3 & 0.1 & 5 & -0.5 & \checkmark & \checkmark & 5.6 (0.4) & 34.2 (3.6) & 6.4 (0.6) & 5.6 (0.4) & 1.2 (0.1) \\
				4 &1.0 & 0.1 & 5 & -0.5 & \checkmark & \checkmark & 20.6 (1.5) & 78.0 (6.0) & 25.1 (2.2) & 20.0 (1.5) & 4.5 (0.7) \\
				5 &2/3 & 0.2 & 5 & -0.5 & \checkmark & \checkmark & 2.8 (0.2) & 31.3 (3.2) & 3.2 (0.3) & 2.9 (0.2) & 0.8 (0.1) \\
				6 &2/3 & 0.1 & 10 & -0.5 & \checkmark & \checkmark & 4.0 (0.3) & 30.8 (2.2) & 4.3 (0.3) & 3.9 (0.3) & 0.8 (0.1)\\
				7 &2/3 & 0.1 & 5 & -0.5 & \checkmark &  &  6.4 (0.5) & 37.4 (2.9) & 7.1 (0.7) & 6.3 (0.5) & 1.7 (0.2) \\
				8 &2/3 & 0.1 & 5 & -0.5 &  & \checkmark & 2.5 (0.2) & 32.8 (4.4) & 3.0 (0.3) & 2.7 (0.3) & 1.6 (0.2) \\
                9 &2/3 & 0.1 & 5 & 0.0 & \checkmark & \checkmark & 3.8 (0.3) & 14.0 (1.5) & 4.0 (0.4) & 3.9 (0.3) & 1.1 (0.1) \\
                10 &2/3 & 0.1 & 5 & 0.5 & \checkmark & \checkmark & 5.9 (0.4) & 14.0 (1.2) & 6.1 (0.5) & 6.2 (0.5) & 1.5 (0.2) \\
			\end{tabular}}
    \vspace{0.1cm}\\
    \text{(b) Mean squared error}
    {\renewcommand{\arraystretch}{1.11}%
    \begin{tabular}{c|cccccc|ccccc}
				\multicolumn{11}{c}{}\\
				
				&$\phi$ & $\rho$ & $\lambda^*$ & $\gamma_u$ & G & S & Naive & Shared & PSA-B & PSA-G & Full\\
				\hline
				1 &0.0 & 0.1 & 5 & -0.5 & \checkmark & \checkmark & 95.0 (0.3) & 27.0 (1.4) & 94.0 (0.4) & 94.5 (0.4) & 96.5 (0.2)\\
				2 &1/3 & 0.1 & 5 & -0.5 & \checkmark & \checkmark & 80.5 (1.1) & 20.5 (1.2) & 81.0 (1.1) & 78.0 (1.2) & 93.0 (0.5)\\
				3 &2/3 & 0.1 & 5 & -0.5 & \checkmark & \checkmark & 48.0 (1.8) & 6.5 (0.4) & 53.0 (1.8) & 50.5 (1.8) & 97.5 (0.2)\\
				4 &1.0 & 0.1 & 5 & -0.5 & \checkmark & \checkmark & 32.5 (1.6) & 8.0 (0.5) & 36.0 (1.6) & 38.5 (1.7) & 94.0 (0.4)\\
				5 &2/3 & 0.2 & 5 & -0.5 & \checkmark & \checkmark & 69.0 (1.5) & 10.0 (0.6) & 64.0 (1.6) & 65.5 (1.6) & 96.5 (0.2)\\
				6 &2/3 & 0.1 & 10 & -0.5 & \checkmark & \checkmark & 42.5 (1.7) & 2.5 (0.2) & 44.0 (1.7) & 47.5 (1.8) & 96.0 (0.3)\\
				7 &2/3 & 0.1 & 5 & -0.5 & \checkmark &  & 49.5 (1.8) & 6.5 (0.4) & 49.5 (1.8) & 51.0 (1.8) & 95.5 (0.3)\\
				8 &2/3 & 0.1 & 5 & -0.5 &  & \checkmark & 80.5 (1.1) & 15.0 (0.9) & 80.5 (1.1) & 81.0 (1.1) & 93.5 (0.4)\\
                9 &2/3 & 0.1 & 5 & 0.0 & \checkmark & \checkmark & 64.5 (1.6) & 16.0 (1.0) & 66.5 (1.6) & 64.0 (1.6) & 97.5 (0.2)\\
                10 &2/3 & 0.1 & 5 & 0.5 & \checkmark & \checkmark & 47.0 (1.8) & 14.0 (0.9) & 49.5 (1.8) & 48.5 (1.8) & 94.5 (0.4)\\
			\end{tabular}}
    \vspace{0.1cm}\\
    \text{(c) 95\% coverage probability}
    \vspace{0.1cm}
    \caption{(a) Bias, (b) mean squared error (MSE) and (c) coverage of 95\% intervals for $\triangle$ are reported by comparing the full model to the naive and shared models, and the outcome regression models with propensity score adjustments. We refer to the adjustment with BART as ``PSA-B", while we denote the adjustment with spatial generalized models as ``PSA-G". The first six scenarios are defined by the strength of preferential sampling ($\phi)$, the strength of spatial dependence $(\rho)$ and the intensity of sampling locations ($\lambda^*$).  The seventh and eighth scenarios generate data with non-Gaussian and non-stationary spatial processes, where ``G" stands for Gaussianity and ``S" stands for stationarity. The final two scenarios represent independence and positive correlation between $Y_0(\bs)$ and $Y_1(\bs)$ at each location $\bs$, signified by the values of $\gamma_u$. All values are multiplied by 100 and standard errors are given in parentheses.}
    \label{t:sim_result_pate}
\end{table}

\section{Analysis of Australian marine protected areas}\label{sec:app}

\subsection{Data description}\label{s:app:des}
\cite{gill2017capacity, gill2024diverse} compiled underwater visual census data from MPA and non-MPA sites worldwide to understand the effect of MPA management and other factors upon the conservation outcomes. For more information on data compilation and the original data sources, see \cite{gill2024diverse}. 
In this analysis, we focus on fish survey sites in Australian waters, comparing the effectiveness of MPA policies with areas that have no protection. The observed outcome $Y(\bs)$ at a location $\bs\in\calD$ is the logarithm of fish biomass density ($\text{g/}100\text{m}^2$) defined in \cite{gill2017capacity}. Out of $3,553$ survey sites, $2,609$ received the marine protection policy ($A(\bs)=1$) and $944$ did not ($A(\bs)=0$). 
\textcolor{black}{\cite{gill2017capacity} assigned policy status to the survey sites based on information from the data provider, time since the establishment of MPAs, and locations relative to the MPA boundaries. See \cite{gill2017capacity} for more information.}
The non-MPA sites within one kilometer of the MPA boundaries were excluded from the original data to avoid a potential interference effect \citep{gill2017capacity, gill2024diverse}, which supports Assumption~\ref{sutva-std}. 

\begin{figure}[t]
\centering
\includegraphics[scale=0.4]{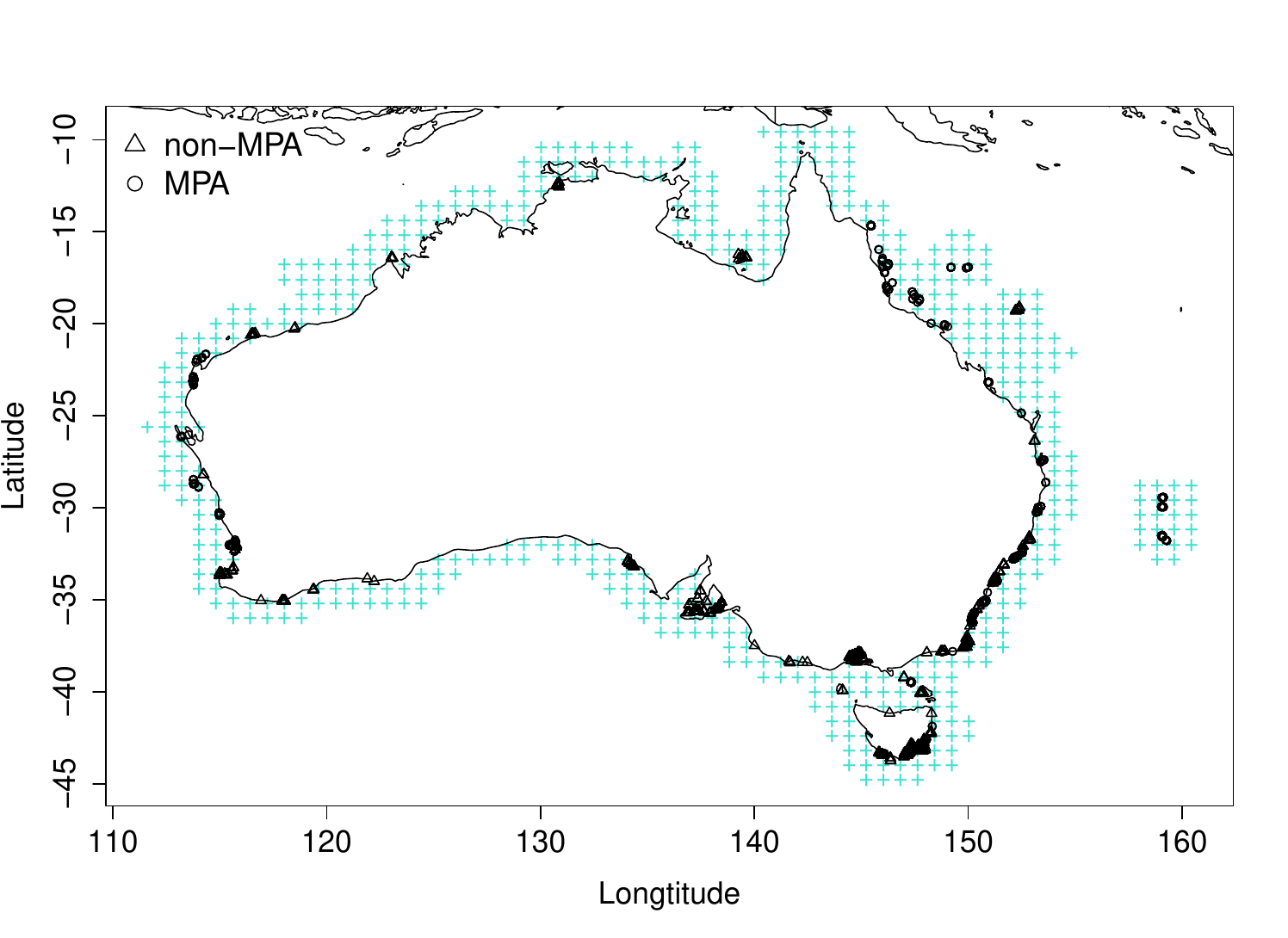}
\caption{Domain of interest (Turquoise) and grid cell locations. Circular dots represent MPA sites, and triangular dots represent non-MPA sites.}
\label{domain}
\end{figure}

We use $p=7$ covariates: distance to shoreline (km), depth (m), mean chlorophyll-a (year 2002 -- 2009, $\text{mg/m}^3$), minimum sea surface temperature (year 2002 -- 2009, $^{\circ}$C), human population density within a radius of 100 km from survey sites (year 2000, the number of individuals), distance to nearest markets (km) and habitat type (coral or rocky reefs). \textcolor{black}{The inclusion of diverse covariates, along with the incorporation of latent processes, is expected to satisfy Assumption~\ref{ignor}.} The spatial domain of interest is confined to the region within $0.5 m_S$ from the shoreline or $150\text{ km}$ within the survey sites, covered by grid cells $0.8 \times 0.8$ $\text{degree}^2$, where $m_S\approx265\text{ km}$ is the maximum distance between the survey sites and the shoreline, approximately, and the number of grid cells is $G=465$ cells (Figure~\ref{domain}). We also fit the model with coarser and finer values of grid cell resolutions and the results are similar (see S5 in Supplemental Material). This domain is constrained to regions where environmental and social factors are similar to the survey sites. Covariates are obtained at the centroids of the grid cells and at the coordinates of the survey sites by following or slightly modifying the procedure in \cite{gill2017capacity} (see S4 in Supplemental Material). For the human population density, the values are log transformed after adding 1 to mitigate skewness. 
Missing values in the mean chlorophyll-a and the minimum sea surface temperature for the grid cells are imputed using the sample means of those with complete data, while missing values for the survey sites are imputed using the mean chlorophyll-a and the minimum sea surface temperature of the grid cells where the survey sites are located. The grid-level covariates and those measured at the survey sites are scaled to collectively have mean zero and variance one.

\subsection{Model details}\label{s:app:model}

We consider both the ``full" model in \eqref{hierar-car1} -- \eqref{hierar-car6} and ``naive" model, which assumes no preferential sampling with $\phi_0 = \phi_1 = 0$. Exploratory analysis suggested similar covariate effects across policies so we set $\bbeta_0 = \bbeta_1 = \bbeta$ and $\bdelta_0 = \bdelta_1 = \bdelta$. \textcolor{black}{Furthermore, we conducted preliminary data analyses to assess for a linear relationship between $\bX(\bs)$ and $Y(\bs)$. Based on a linear regression model, we did not detect any significant nonlinearities of the residuals.} We assign the same prior distributions as those defined in Section~\ref{s:sim:dgp} to most model parameters, except for the spatial dependence parameters, which we assign $\log(\rho_u)$ and $\log(\rho_v)$ each $\text{Normal}\{\log(300), 0.5\}$, and $\log(\kappa_u)$ and $\log(\kappa_v)$ each $\text{Normal}\{\log(0.5), 1\}$. MCMC is again used for model inference where the chains are run for $205,000$ iterations and the first 5,000 are discarded as burn-in. The resulting chain is thinned to every 10th such that $20,000$ posterior samples are retained for inference.

\subsection{Results}\label{s:app:results}

We compare the full and naive models to investigate the impacts of preferential sampling on model inference, including estimating causal effects. Table~\ref{post.mean.beta.delta} summarizes the posterior distributions of $\bbeta$ and $\bdelta$ in the full model. The depth, minimum sea surface temperature, and distance to the markets have statistically significant effects. Fish biomass tends to increase as the survey site is deeper, which coincides with research that bathymetry and fish biomass are positively correlated \citep{boswell2010biomass}. More fish biomass is expected as the sea surface temperature increases. We conjecture that this is due to a negative correlation between sea surface temperature and habitat type. Coral reefs tend to be in the northern part of Australia, where sea temperature is relatively higher than in the southern part with more rocky reefs. We obtain a positive posterior mean estimate of the coefficient of distance to markets, which is natural since fish biomass is higher in regions with less human accessibility \citep{williams2008assessing}. The depth and the human population density have statistically significant effects on the sampling process. The negative coefficient estimate of depth and the positive coefficient estimate of human population density indicate that the survey tended to be focused on sites with greater accessibility.

\begin{table}[t]
    \centering
    \resizebox{\columnwidth}{!}{%
        \begin{tabular}{c|ccccccc}
             & shoreline & depth & chlorophyll & temperature & population & market & habitat \\
             \hline
            $\bbeta$ & -0.08 (0.13) & 9.25 (1.81) & 0.06 (0.04) & 1.01 (0.17) & -0.08 (0.24) & 0.63 (0.27) & -0.76 (0.63)\\
            $\bdelta$ & -0.95 (0.54) & -1.03 (0.29) & -0.44 (0.85) & -1.02 (1.07) & 2.36 (0.67) & 0.06 (0.73) & -0.56 (1.99) \\
    \end{tabular}
    }
    \caption{Posterior means (standard deviations) of the covariate effects on biomass ($\bbeta$) and sampling intensity ($\bdelta$) for the full model applied to the Australia coast. }
    \label{post.mean.beta.delta}
\end{table}


Table~\ref{post.mean.others} summarizes the posterior distributions of the parameters $\alpha_0$, $\alpha_1$, $\phi_0$, and $\phi_1$ from the full model. Since the covariates are mean-centered, the posterior means of $\alpha_1$ and $\alpha_0$ can be interpreted as the average fish biomass with and without MPA policies respectively, when all other covariates are set to their mean values. The difference $\alpha_1-\alpha_0$ can thus be interpreted as a global causal effect, which has a posterior mean $0.61$ and standard deviation $0.74$. This implies that the implementation of MPAs has a positive effect on the preservation of fish biodiversity, although it is not significant.

The positive posterior mean of $\phi_0$ reported in Table~\ref{post.mean.others} suggests non-MPA sites are sampled where fish biomass is high, while the negative mean of $\phi_1$ indicates MPA sites are sampled where biomass is lower. We conjecture that the survey was conducted for MPA sites in favor of more accessible regions in which fish biomass is lower due to human access. Evidence can be found from the posterior mean values of $\bdelta$ in Table~\ref{post.mean.beta.delta}, suggesting that the surveys were carried out in regions closer to the shorelines, with shallower depth and with a larger human population. The posterior mean and the 95\% credible interval of $\phi_1 - \phi_0$ are $-1.70$ and $(-5.67, 1.26)$. Since approximately 82\% of the posterior mass of $\phi_1 - \phi_0$ is negative, we suggest that the degree of preferential sampling varies between the MPA and non-MPA sites.

\begin{table}[t]
    \centering
    \resizebox{\columnwidth}{!}{%
        \begin{tabular}{c|cccccccc}
             & $\alpha_0$ & $\alpha_1$ & $\phi_0$ & $\phi_1$ & $\gamma_u$ & $\gamma_v$ & $r_u$ & $r_v$  \\
             \hline
            Mean (SD) & 7.24 (1.04) & 7.85 (1.14) & 0.61 (1.34) & -1.09 (2.17) & 0.44 (0.33) & 5.40 (6.22) & 0.46 (0.32) & 0.70 (0.47)
        \end{tabular}
    }
    \caption{Posterior means (standard deviations) of parameters for the full model}
    \label{post.mean.others}
\end{table}

Table~\ref{post.mean.others} also summarizes the posterior distributions of the parameters $\gamma_u$, $\gamma_v$, $r_u$, and $r_v$ from the full model, where $r_u$ (and similarly $r_v$) is defined as the correlation coefficient between $U_{0,g}$ and $U_{1,g}$ for all $g$, computed as
\begin{equation}
    r_u = \frac{\gamma_u\cdot\sigma_{u,0}^2}{\sqrt{\sigma_{u,0}^2(\sigma_{u,1}^2+\gamma_u^2\cdot\sigma_{u,0}^2})}.
\end{equation}
The positive posterior means of $r_u$ and $r_v$ suggest that the spatial processes $U_{0,g}$ and $U_{1,g}$ are positively correlated, and so are $V_{0,g}$ and $V_{1,g}$.

\begin{figure}[t]
    \centering
    \includegraphics[scale=0.4]{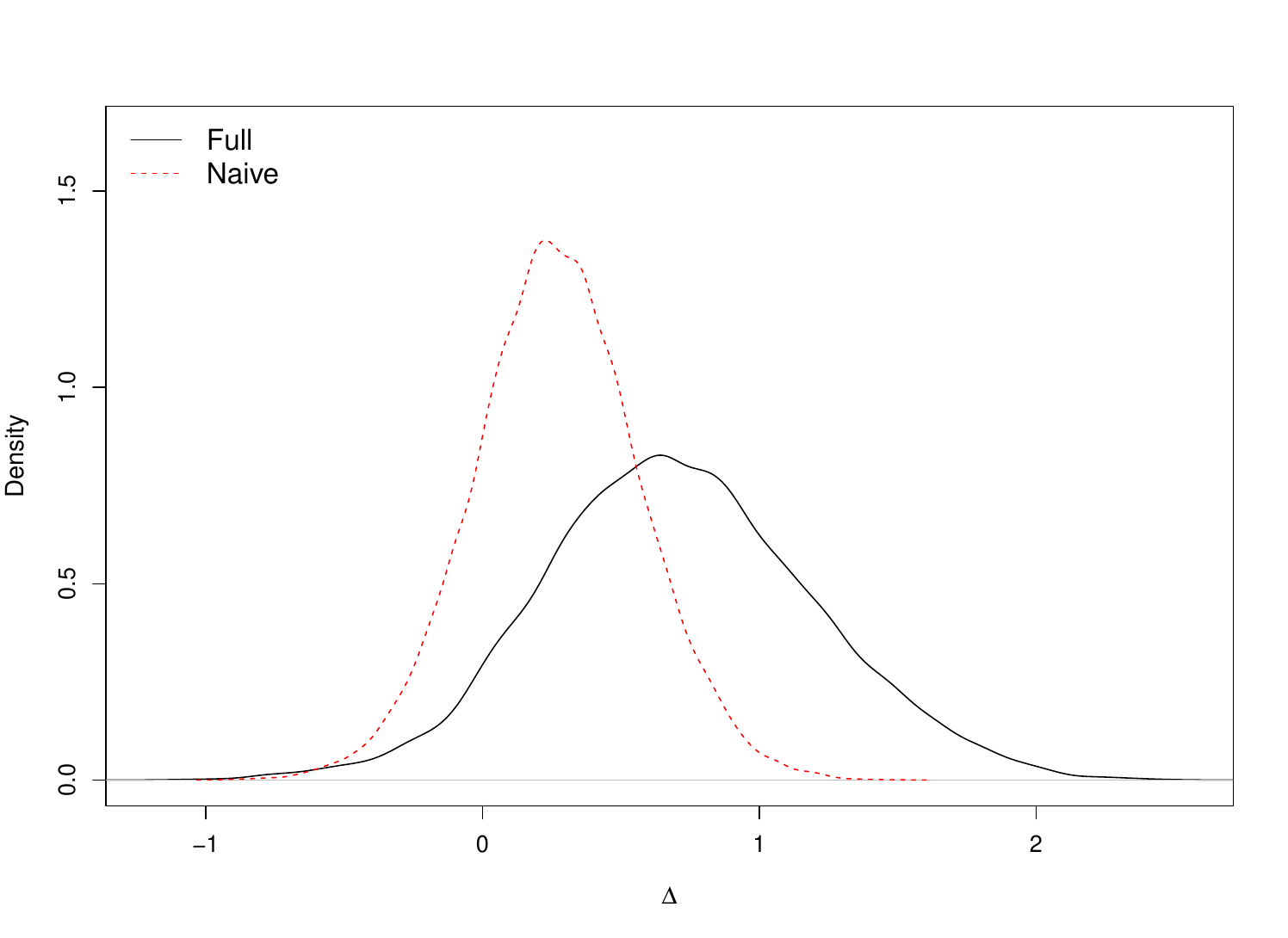}   
    \caption{Posterior distributions of $\triangle$ (logarithm of $\text{g/}100\text{m}^2$) under the full and naive models.}
    \label{PATE-fish}
\end{figure}

The marginal posterior distribution of $\triangle$ under both models is shown in Figure~\ref{PATE-fish}. The mean posteriors of $\triangle$ from the full and naive models are $0.71$ and $0.27$, respectively. Although both models estimate that implementing MPAs is superior to not doing so in terms of fish biodiversity on average, the naive model produces more conservative results compared to the full model. The 95\% credible intervals being $(-0.24, 1.71)$ and $(-0.32, 0.87)$ under the full and naive models, respectively.

\begin{figure}
    \centering
    \includegraphics[scale=0.48]{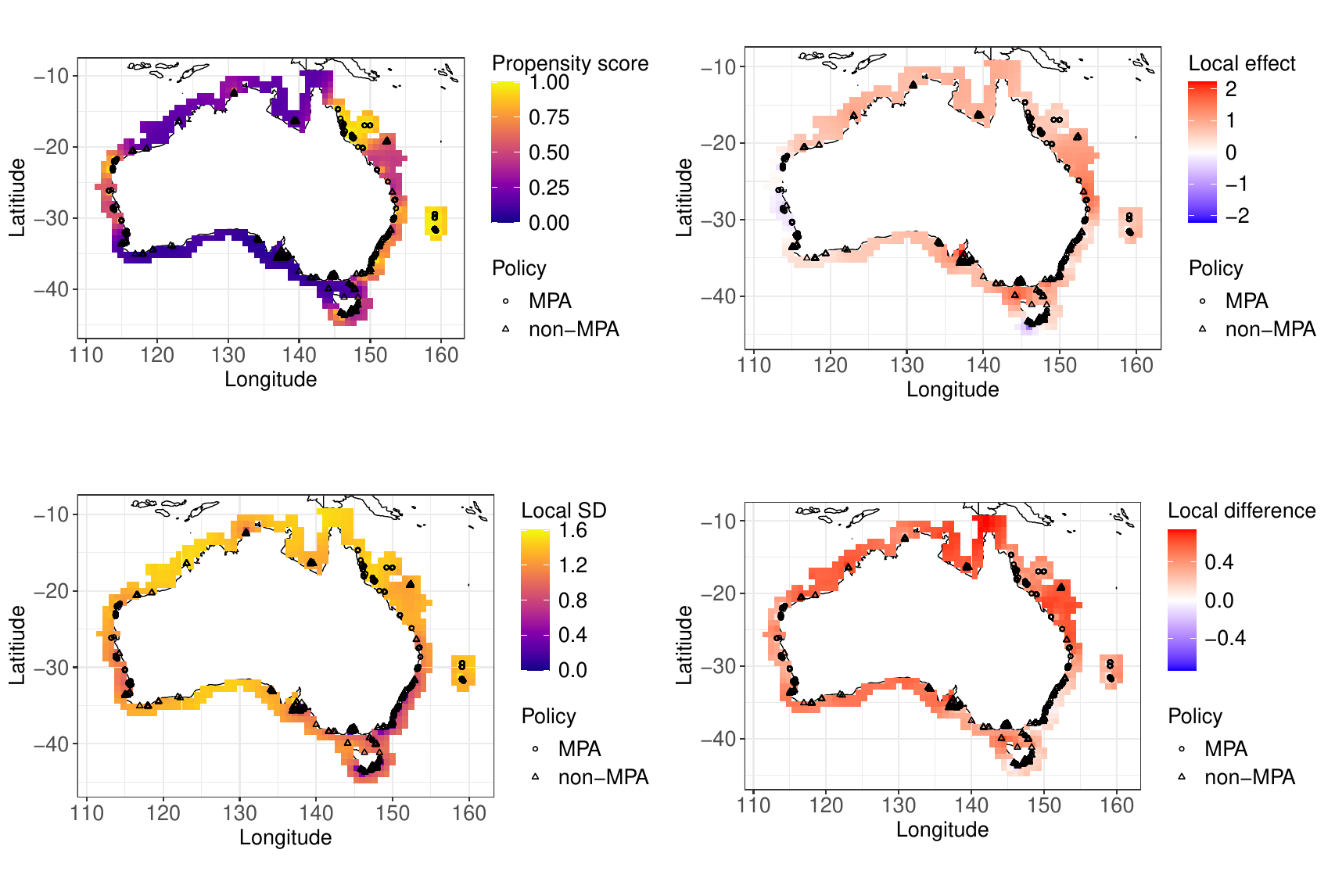}
    \caption{Posterior means of propensity scores $\frac{\lambda_{1,g}}{\lambda_{0,g}+\lambda_{1,g}}$  at each grid cell, along with the sampling locations (Upper left), posterior mean (Upper right) and standard deviation (Lower left) of local causal effects $\triangle_g$ under the full model, and the difference between posterior mean of local causal effects $\triangle_g$ under the full and naive model (Lower right)}
    \label{all_plots}
\end{figure}

The upper left panel of Figure~\ref{all_plots} presents the posterior means of the propensity score $\frac{\lambda_{1,g}}{\lambda_{0,g} + \lambda_{1,g}}$ at each grid cell. \textcolor{black}{The propensity scores are estimated to be strictly between 0 and 1 for most regions, verifying Assumption~\ref{posi_int}}. The propensity score estimate in the northern and southern parts of $\calD$ indicate that these regions are less likely to be assigned MPA policies, whereas the estimates in the eastern and western parts of $\calD$ are more likely to be assigned MPA policies. Specifically, the eastern and western regions of Australia align with the coral reef area (\citealp[Figure 1 of ][]{dixon2022coral}), where their conservation is of great concern \citep{edgar2014global}. This is potential evidence that the MPA sites were sampled preferentially. 

The upper right and lower left panels of Figure~\ref{all_plots} show the posterior means and standard deviations, respectively, of the local causal effects $\alpha_1-\alpha_0 + U_{1,g} - U_{0,g}$. Local causal effects are positive for the majority of the grid cells, implying that the implementation of MPAs is helpful in preserving fish biodiversity in these grid cells. The lower right panel of Figure~\ref{all_plots} presents the difference between the posterior mean of the local causal effect under the full and naive model. As indicated in the figure provided, the naive model tends to give smaller estimates of the local causal effect across most grid cells.

\section{Discussion}\label{s:discussion}
In this article, we present a novel approach to spatial causal inference that adjusts for bias introduced by preferential sampling. We model the response processes and point processes jointly using shared random effects to address preferential sampling. Our theoretical studies reveal that the parameters of interest are identifiable and the posterior distribution of the model parameters is weakly consistent. Our simulation study shows that the proposed model has valid coverage under the scenarios we considered, and that both bias and mean-squared error are reduced when compared to the naive model that does not account for preferential sampling. We apply the proposed method to study the impacts of MPA implementation on fish biomass on the Australian coast. We find evidence of preferential sampling and that consideration of the preferential sampling affects the causal effect.

To examine the plausibility of the SUTVA, latent ignorability, and positivity assumptions for the application to MPA data, we argue that the SUTVA assumption is justifiable since the non-MPA sites within one kilometer of the MPA boundaries were removed from the data to prevent spatial interference \citep{gill2017capacity, gill2024diverse}. Although latent ignorability is typically not a testable assumption, similar to the unconfoundedness assumption in classical causal inference \citep{li2023bayesian}, we include many key covariates, \textcolor{black}{as well as the shared latent spatial confounder, to support this assumption}. Finally, we suggest that the positivity assumption is satisfied as the posterior means of the propensity scores are strictly between 0 and 1.

Some limitations of the proposed model exist that can be improved in future research. First, even if some non-MPA sites showing proximity to MPA regions were excluded from the data, there can be potential interference between MPA and non-MPA sites due to various attributes such as fish movement. Second, the proposed model specifies the mean of the response variables and the log intensities of the sampling process as linear models, which may be unrealistic. Third, the proposed model assumes that the propensity scores of all units in a grid cell are the same \textcolor{black}{even though policies (treatments) are assigned at the location level. Furthermore, the proposed method postulates the outcome regression models to adjust for the unmeasured spatial confounders. However, the method does not directly model the propensity scores. We speculate that this creates imbalances in the estimated propensity scores between the policy groups at certain values. See the supporting information for more details regarding the propensity score overlap.} Future work is required for additional flexibility. \textcolor{black}{Fourth, while we assume binary treatments, continuous exposures have been of recent interest in the causal inference literature. Joint modeling of the point processes and outcome regressions with continuous treatments could be considered in future work, motivated by \cite{kennedy2017non} and \cite{ woody2020estimating}.} Finally, sensitivity analysis is necessary to examine whether there exist unmeasured confounders, since social, political, and ecological factors determine MPA placement \citep{devillers2015reinventing}, sampling of survey points, and fish biomass.
Considering the aspects mentioned above for developing spatial causal inference methods taking the preferential sampling into account represents a promising direction for future research.

\begin{funding}
This work was supported in part by NSF grant DMS2152887, NIH grants R01ES031651-01 and  1R01ES036270-01A1, and USGS Climate Adaptation Science Center grant G22AC00597-01.
\end{funding}


\begin{supplement}
\stitle{Supplemental Material}
\sdescription{The supplemental material includes the derivation of \eqref{e:bias}, the proofs for Theorem~\ref{identifiability} and \ref{weak-consistency}, the derivation of Gibbs samplers, the data source and the analysis of sensitivity to the grid cell resolutions.}
\end{supplement}

\begin{supplement}
\stitle{Code}
\sdescription{The R code for the proposed method and its competitors is provided in this Supplementary Material and is also available in the GitHub repository \url{https://github.com/Dongjae-Son/SpatCausal.git}, along with the corresponding data.}
\end{supplement}


    \bibliographystyle{imsart-nameyear} 
    \bibliography{refs}

\end{document}


\def\spacingset#1{\renewcommand{\baselinestretch}%
{#1}\small\normalsize} \spacingset{1}

\thispagestyle{empty}
\begin{center}
{\Large Supplemental Material of: \\
Spatial causal inference in the presence of preferential sampling to study the impacts of marine protected areas}\\\vspace{6pt}
{Dongjae Son, Brian J. Reich, Erin M. Schliep, Shu Yang, David A. Gill}\\\vspace{6pt}
\end{center}


\renewcommand{\thesubsection}{S\arabic{subsection}}
\spacingset{1.9} 
\subsection{Details of the bias calculation}\label{s:A1}

In this section we derive the approximate bias introduced to the causal effect by preferential sampling given in the expression (11). We let observed covariates $\bX(\bs)$ for $\bs\in\calD$ fixed. By the law of iterated expectations, we have
\begin{equation*}
    \mathbb{E}\left\{Y_1(\bs)\right\} = \mathbb{E}_{\calW}\left[\mathbb{E}_\calS\left\{\alpha_1 + \bX(\bs)^{\top}\bbeta_1+U_1(\bs)\right\}\right]
\end{equation*}
where $\mathbb{E}_\calW(\cdot)$ and $\mathbb{E}_\calS(\cdot)$ are expectations with respect to spatial random effects and point processes respectively. Then,
\begin{align*}
    &\mathbb{E}_\calS\left\{\alpha_1 + \bX(\bs)^{\top}\bbeta_1+U_1(\bs)\right\} \\
    &= \alpha_1 + \frac{\int_\calD \left\{\bX(\bw)^{\top}\bbeta_1 + U_1(\bw)\right\} \cdot \exp\left\{\bX(\bw)^{\top}\bdelta_1 + V_1(\bw) + \phi_1U_1(\bw)\right\}d\bw}{\int_\calD \exp\left\{\bX(\bw)^{\top}\bdelta_1 + V_1(\bw) + \phi_1U_1(\bw)\right\}d\bw}
\end{align*}
since the location densities have form (6). Then $\mathbb{E}\left\{Y_1(\bs)\right\}$ is equivalent to
\begin{equation*}
    \alpha_1 + \mathbb{E}_{\calW}\left[\frac{\int_\calD \left\{\bX(\bw)^{\top}\bbeta_1 + U_1(\bw)\right\} \cdot \exp\left\{\bX(\bw)^{\top}\bdelta_1 + V_1(\bw) + \phi_1U_1(\bw)\right\}d\bw}{\int_\calD \exp\left\{\bX(\bw)^{\top}\bdelta_1 + V_1(\bw) + \phi_1U_1(\bw)\right\}d\bw}\right].
\end{equation*}
Using the first-order Taylor expansion, the expectation of the ratio is approximated to the ratio of the expectations, i.e.,
\begin{equation*}
    \mathbb{E}\left\{Y_1(\bs)\right\}\approx\alpha_1 + \frac{\mathbb{E}_{\calW}\left[\int_\calD \left\{\bX(\bw)^{\top}\bbeta_1 + U_1(\bw)\right\} \cdot \exp\left\{\bX(\bw)^{\top}\bdelta_1 + V_1(\bw) + \phi_1U_1(\bw)\right\}d\bw\right]}{\mathbb{E}_{\calW}\left[\int_\calD \exp\left\{\bX(\bw)^{\top}\bdelta_1 + V_1(\bw) + \phi_1U_1(\bw)\right\}d\bw\right]}.
\end{equation*}
By Fubini's theorem, we can interchange the integral and the expectation in the denominator of the second term, i.e.,
\begin{align*}
    \int_\calD \mathbb{E}_{\calW}\left[\exp\left\{\bX(\bw)^{\top}\bdelta_1 + V_1(\bw) + \phi_1U_1(\bw)\right\}\right]d\bw.
\end{align*}
Note that $\bX(\bw)^{\top}\bdelta_1 + V_1(\bw) + \phi_1U_1(\bw)$ follows a normal distribution with mean $\bX(\bw)^{\top}\bdelta_1$ and variance $\Var\left(V_1\right) + \phi_1^2 \Var\left(U_1\right)$. The variance does not depend on $\bw$ since we assume stationary processes. Accordingly, we have
\begin{align*}
    \int_\calD\mathbb{E}_{\calW}&\left[\exp\left\{\bX(\bw)^{\top}\bdelta_1 + V_1(\bw) + \phi_1U_1(\bw)\right\}\right]d\bw \\
    &= \exp\left\{\frac{\Var\left(V_1\right) + \phi_1^2 \Var\left(U_1\right)}{2}\right\}\int_\calD\exp\left\{\bX(\bw)^{\top}\bdelta_1\right\}d\bw.
\end{align*}
Using the law of iterated expectations, we also have
\begin{align*}
    \int_\calD\mathbb{E}_{\calW}&\left[U_1(\bw)\cdot\exp\left\{\bX(\bw)^{\top}\bdelta_1 + V_1(\bw) + \phi_1U_1(\bw)\right\}\right]d\bw \\
    &=\phi_1\Var(U_1)\cdot\exp\left\{\frac{\Var\left(V_1\right) + \phi_1^2 \Var\left(U_1\right)}{2}\right\}\int_\calD\exp\left\{\bX(\bw)^{\top}\bdelta_1\right\}d\bw.
\end{align*}
Therefore, we have $\mathbb{E}\left\{Y_1(\bs)\right\} \approx \alpha_1 + {\bar \bX}_1\bbeta_1 +\phi_1\Var(U_1)$. If we do the same steps for $\mathbb{E}\left\{Y_0(\bs)\right\}$, the expectation becomes (11).

\subsection{Proofs of Theorem 1 and Theorem 2}\label{s:A3}

\vspace{10pt}
\noindent
{\bf The proof of the Theorem 3.1}

\noindent
Our final model discretizes the spatial domain $\calD$ and assigns observations to $G$ equally-sized grid cells $\{\calD_1,...,\calD_G\}$ with centroids $\{\bc_1,...,\bc_G\}$.  Recall that $g(\bs_i)$ be the index of the grid cell assigned to the observation at $\bs_i$, i.e., $g(\bs_i) = k$ if $\bs_i\in\calD_k$. The covariates and spatial processes at the centroids are denoted $\bX(\bc_k) = \bX_k$, $U_{a}(\bc_k) = U_{a,k}$ and $V_{a}(\bc_k) = V_{a,k}$. The joint model for $a\in\left\{0,1\right\}$ is
\begin{align*}
    &Y_{a}(\bs_i)|\bs_i, \bX(\bs_i), \btheta, \calW \overset{\text{indep}}{\sim}\text{Normal}\left\{\alpha_a + \bX(\bs_i)^{\top}\bbeta_a + U_{a,g(\bs_i)}, \tau_a^2\right\}  \\
    &p_a\left(\bs_i\in\calD_k|\calX, \btheta, \calW\right)=\pi_{a,k} = \frac{\exp\left\{\bX_k^{\top}\bdelta_a + {V}_{a,k} + \phi_aU_{a,k}\right\}}{\sum_{g=1}^G \exp\left\{\bX_g^{\top}\bdelta_a + {V}_{a,g} + \phi_aU_{a,g}\right\}} \\
    &(U_{a,1},...,U_{a,G})^\top|\btheta \sim \text{Normal}\{\mathbf{0}_G, \bR_{U_a}(\btheta_C)\} \\
    &(V_{a,1},...,V_{a,G})^\top|\btheta \sim \text{Normal}\{\mathbf{0}_G, \bR_{V_a}(\btheta_C)\}
\end{align*}
where ${\calW} = \left\{\left(\Tilde{U}_a(\bs), \Tilde{V}_a(\bs)\right):\bs\in\calD, a\in\left\{0,1\right\}\right\}$, and the covariance matrices $\bR_{U_a}$ and $\bR_{V_a}$ are defined by the parameters $\btheta_C$. 

To show identifiability using only the data requires marginalizing over the latent processes $\calW$, which is immediate for the outcome variables but challenging for the Poisson process for the sampling locations. However, under the positivity assumption, for large $n$ the proportion of the observations that fall in grid cell $k$ will converge to $\pi_{ak} = \lambda_{a,k}/\sum_{g=1}^G\lambda_{a,g}>0$. 
The set of $\pi_{ak}$ is one-to-one with the set of linear predictors $L_{a,k} = \log\left(\lambda_{a,k}\right) = \eta_a + \bX_k^{\top}\bdelta_a + {V}_{a,k} + \phi_aU_{a,k}$ across the grid cells, which follows a multivariate normal distribution, facilitating marginalization over $\calW$. Then the marginal distributions of $\bY_a = (Y_{a}(\bs),...,Y_{a}(\bs_n))^{\top}$ and $\bL_{a} = (L_{a,1},...,Z_{a,G})^{\top}$ are 
\begin{align*}
    &\bY_{a}|\btheta \sim\text{Normal}\left\{\alpha_a\mathbf{1}_n + \bX_\calS\bbeta_a, \bH\hspace{1pt}\bR_{U,a}(\btheta_C)\hspace{1pt}\bH^{\top} + \tau_a^2\mathbb{I}_n \right\} \\
    &\bL_{a}|\btheta \sim\text{Normal}\left\{\eta_a\mathbf{1}_G + \bX_\calG\bdelta_a, \bR_{V_a}(\btheta_C) + \phi_a^2\bR_{U_a}(\btheta_C) \right\}
\end{align*}
where $\bX_\calS$, $\bH$, and $\bX_\calG$ are the same as defined in Appendix~\ref{s:A1}.
      
We say that $\btheta$ is identifiable if and only if $\ell\left(\bD|\btheta^{(1)}\right)=\ell\left(\bD|\btheta^{(2)}\right)$ implies $\btheta^{(1)} = \btheta^{(2)}$ where $\ell()$ is a generic notation for likelihood and $\bD=(\bY_0, \bY_1, \bL_0, \bL_1)$. Since both $\bY_a$ and $\bL_a$ are normally distributed, so is $\bD$. Then, showing the identifiability is equivalent to proving that the mean vector and covariance matrix of $\bD$ parametrized by $\btheta$ is an one-to-one function of $\btheta$.


We start with the identification of $\phi_0$. Comparing $\bs$ and $\bs_{i'}$ in different grid cells immediately identifies $\phi_0$ since
\begin{align*}
    &\text{Cov}\left\{Y_0(\bs_i), Y_{0}(\bs_{i'})\right\} = \sigma_{u,0}^2\mathcal{R}\left(\|\bc_{g(\bs_i)}-\bc_{g(\bs_{i'})}\|;\kappa_u, \rho_u\right) \text{ and } \\
    &\text{Cov}\left\{Y_0(\bs_i), L_{0,g(\bs_{i'})}\right\} = \phi_0\sigma_{u,0}^2\mathcal{R}\left(\|\bc_{g(\bs_i)}-\bc_{g(\bs_{i'})}\|;\kappa_u, \rho_u\right).
\end{align*}
Furthermore, since $\Cov\left\{Y_0(\bs_i), L_{0,g(\bs_i)}\right\} = \phi_0\sigma_{u,0}^2$ and since $\phi_0$ has been identified, so has $\sigma_{u,0}^2$. To identify $\tau_0^2$, $\kappa_u$, and $\rho_u$, suppose $\text{Var}\left(\bY_0|\btheta^{(1)}\right)=\text{Var}\left(\bY_0|\btheta^{(2)}\right)$. Comparing the diagonal elements is equivalent to $\sigma_{u,0}^2 + \left(\tau_0^{(1)}\right)^2 = \sigma_{u,0}^2 + \left(\tau_0^{(2)}\right)^2$, implying that $\left(\tau_0^{(1)}\right)^2 = \left(\tau_0^{(2)}\right)^2$. Thus, $\tau_0^2$ is identified. Furthermore, comparing the off-diagonal elements is equivalent to
\begin{equation*}
    \sigma_{u,0}^2\mathcal{R}\left(\|\bc_{g(\bs_i)}-\bc_{g(\bs_{i'})}\|;\kappa_u^{(1)},\rho_u^{(1)}\right)=\sigma_{u,0}^2\mathcal{R}\left(\|\bc_{g(\bs_i)}-\bc_{g(\bs_{i'})}\|;\kappa_u^{(2)},\rho_u^{(2)}\right)
\end{equation*}
for all $i,i'\in\left\{1,\cdots,n\right\}$ since we have $\frac{n(n-1)}{2}$ equations for $2$ variables. Therefore, $\kappa_u$ and $\rho_u$ are identified. See \cite{de2022information} for more details regarding estimation of the parameters of the Matérn covariance function. They suggested that geostatistical data contain as much information for the estimation of the smoothness parameter of a Matérn kernel as the range parameter, which is opposed to the tradition that rather fixes the smoothness parameter.

For the identification of $\gamma_u$, observe that $\Cov\left\{Y_0(\bs_i), Y_1(\bs_i)\right\} = \gamma_u\sigma_{u,0}^2$. Since $\sigma_{u,0}^2$ has been identified, $\gamma_u$ is also identified. Now suppose $\Var\left(\bY_1|\btheta^{(1)}\right)=\Var\left(\bY_1|\btheta^{(2)}\right)$. By comparing the off-diagonal elements, we have 
\begin{equation*}
    \left\{\left(\sigma_{u,1}^{(1)}\right)^2 + \gamma_u^2\sigma_{u,0}^2\right\}\mathcal{R}\left(\|\bc_{g(\bs_i)}-\bc_{g(\bs_{i'})}\|;\kappa_u,\rho_u\right) = \left\{\left(\sigma_{u,1}^{(2)}\right)^2 + \gamma_u^2\sigma_{u,0}^2\right\}\mathcal{R}\left(\|\bc_{g(\bs_i)}-\bc_{g(\bs_{i'})}\|;\kappa_u,\rho_u\right)
\end{equation*} 
implying that $\left\{\sigma_{u,1}^{(1)}\right\}^2=\left\{\sigma_{u,1}^{(2)}\right\}^2$. Therefore, $\sigma_{u,1}^2$ is identified. Then, by comapring the diagonal elements from $\Var\left(\bY_1|\btheta^{(1)}\right)=\Var\left(\bY_1|\btheta^{(2)}\right)$, we can identify $\tau_1^2$. Since $\Var(L_{0,g}) = \sigma_{v,0}^2 + \phi_0^2\sigma_{u,0}^2$, comparing the diagonal elements of $\Var\left(\bL_0|\btheta^{(1)}\right)=\Var\left(\bL_0|\btheta^{(2)}\right)$ immediately identifies $\sigma_{v,0}^2$. Similarly as above, comparing the off-diagonal elements yields
\begin{equation*}
    \sigma_{v,0}^2\mathcal{R}\left(\|\bc_{g}-\bc_{g'}\|;\kappa_v^{(1)},\rho_v^{(1)}\right)=\sigma_{v,0}^2\mathcal{R}\left(\|\bc_{g}-\bc_{g'}\|;\kappa_v^{(2)},\rho_v^{(2)}\right)
\end{equation*}
for all $g,g'\left\{1,\cdots,G\right\}$, implying that $\kappa_v$ and $\rho_v$ are identified. 

To identify $\phi_1$, consider $$\Cov\left\{Y_1(\bs_i), L_{1,g(\bs_i)}\right\} = \Cov\left\{U_{1,g(\bs_i)}, \phi_1U_{1,g(\bs_i)}\right\} = \phi_1\left(\sigma_{u,1}^2+\gamma_u\sigma_{u,0}^2\right).$$
Since $\sigma_{u,0}^2$, $\sigma_{u,1}^2$, and $\gamma_u$ have been identified, the identification of $\phi_1$ is immediate. For identifying $\gamma_v$, consider $\Cov\left(L_{0,g}, L_{1,g}\right) = \gamma_v\sigma_{v,0}^2 + \gamma_u\phi_0\phi_1\sigma_{u,0}^2$. This quantity identifies $\gamma_v$ since $\sigma_{v,0}^2$, $\gamma_u$, $\phi_0$, $\phi_1$, and $\sigma_{u,0}^2$ have been identified. Since $\Var\left(L_{1,g}\right) = \sigma_{v,1}^2 + \gamma_v^2\sigma_{v,0}^2 + \phi_1^2\left(\sigma_{u,1}^2+\gamma_u\sigma_{u,0}^2\right)$, $\sigma_{v,1}^2$ is identified as all the other parameters involving $\Var\left(L_{1,g}\right)$ have been identified. Finally, the identification of $\alpha_a$, $\beta_a$, $\eta_a$ and $\delta_a$ is guaranteed by Assumption 5. This ends the proof of the identifiability of $\btheta$.

\vspace{10pt}
\noindent
{\bf The proof of the Theorem 3.2}

\noindent
Let $\bomega = ({\btheta}_\calM, \calW)$ be the parameters in the causal effect $\triangle$. Denote an arbitrary observation as $\bZ = (Y, A, \bs)$ and note that the observations $(\bZ_1,\cdots,\bZ_n)$ are independently and identically distributed conditioned on $\bomega$. Denote the joint density of $\bZ$ given a set of covariates $\calX$ and parameters $\bomega$ as $f(\bZ|\bomega)$ and the prior of $\bomega$ as
 $\Pi(\bomega)$, where the vector of hyperparameters $\btheta_\calC$ is assigned by the priors that satisfies Assumption 4. To prove weak consistency of a posterior distribution of $\bomega$, we define a Kullback-Leibler (K-L) support \citep{ghosal2017fundamentals, ghosh2003nonpara} to utilize a result from \cite{schwartz1965bayes} that if Definition~\ref{klsupport} is satisfied, then the posterior of $\bomega$ is weakly consistent \citep{ghosh2003nonpara}. 
\begin{definition}\label{klsupport}
    The density $f$ at the true value $\bomega^*$ is said to be in the K-L support of the prior $\Pi$ if
    \begin{equation*}
        \Pi\left\{\bomega: K\left(\bomega^*, \bomega\right) <\varepsilon \right\} >0
    \end{equation*}
    for all $\varepsilon>0$, where $K\left(\bomega^*, \bomega\right) = \int \log\frac{f(\bZ|\bomega^*)}{f(\bZ|\bomega)}f(\bZ|\bomega^*)d\bz$ is a Kullback-Leibler (K-L) divergence.
\end{definition} 
\noindent
To show that $f(\bZ|\bomega^*)$ is in the K-L support of the prior $\Pi$, we define a $\nu$-ball $B_{\nu}(\bomega^*)$ to be
\begin{equation*}
\begin{aligned}[b]
 \cap_{a=0}^1&\left\{\bomega:|\alpha_{a}-\alpha_{a}^*| < \nu,\|\bbeta_{a}-\bbeta_{a}^*\|_2 < \nu, |\eta_{a}-\eta_{a}^*| < \nu, \|\bdelta_{a}-\bdelta_{a}^*\|_2 < \nu, |\phi_a - \phi_{a}^*| < \nu, \right. \\ &|\tau_a / \tau_{a}^* - 1| < \nu, 
\left.\|\Tilde{U}_a - \Tilde{U}_a^*\|_\infty < \nu, \|\Tilde{V}_a - \Tilde{V}_a^*\|_\infty < \nu, |\gamma_u-\gamma_{u}^*|<\nu, |\gamma_v-\gamma_{v}^*|<\nu \right\}
\end{aligned}
\end{equation*}
for some $\nu>0$. The proof proceeds by first showing that for every $\varepsilon>0$ there exists a $\nu>0$ such that for all $\bomega \in B_\nu(\bomega*)$  we have $K(\bomega,\bomega^*) < \varepsilon$. We then show that the prior probability on $\left\{\bomega : \bomega\in B_\nu(\bomega^*)\right\}$ is positive to complete the proof.

Under Assumption 1 and Assumption 2, the joint density can be decomposed as
\begin{equation*}
    f(\bZ|\bomega) = f_a(Y_a|\bs, \bomega)p_a(\bs|\bomega)\mathbb{P}(A=a|\bomega)
\end{equation*}
where $f_a(Y_a|\bs, \bomega)$ is the density of the potential outcome given $\bs$ and $\bomega$, $p_a(\bs|\bomega)$ is the location density for $\calS_a$ given $\bomega$, and $$\mathbb{P}(A=a|\bomega) = \frac{\int_\calD\lambda_a(\bs)d\bs}{\int_\calD\lambda_0(\bs)d\bs+\int_\calD\lambda_1(\bs)d\bs}$$ for $a\in\left\{0,1\right\}$. 

The KL divergence is decomposed into two parts as
\begin{equation*}
    K(\bomega^*, \bomega) = K_1(\bomega^*, \bomega)\mathbb{P}(A=1|\bomega^*) + K_0(\bomega^*, \bomega)\mathbb{P}(A=0|\bomega^*) < K_1(\bomega^*, \bomega) + K_0(\bomega^*, \bomega)
\end{equation*}
where
\begin{equation*}
    K_a(\bomega^*, \bomega) = \int \log\frac{f_a(Y_a|\bs, \bomega^*)p_a(\bs|\bomega^*)\mathbb{P}(A=a|\bomega^*)}{f_a(Y_a|\bs, \bomega)p_a(\bs|\bomega)\mathbb{P}(A=a|\bomega)}f_a(Y_a|\bs, \bomega)p_a(\bs|\bomega)dY_ad\bs.
\end{equation*}

We focus on $K_1(\bomega^*, \bomega)$, and a similar argument can be applied to $K_0(\bomega^*, \bomega)$. The steps given below are based on the proof of the Theorem 1 in \cite{pati2011bayesian}. 
Let $\mu_{1}(\bs) = \alpha_1+ \bX(\bs)^{\top}\bbeta_1 + U_1(\bs)$, $U_1(\bs) = \Tilde{U}_1(\bs) + \gamma_u\Tilde{U}_0(\bs)$ and $V_1(\bs) = \Tilde{V}_1(\bs) + \gamma_v\Tilde{V}_0(\bs)$, while $\mu_{1}^*(\bs)$, $U_{1}^*(\bs)$, and $V_{1}^*(\bs)$ are those evaluated at $\bomega^*$. Then,
\begin{align}
    K_1\left(\bomega^*,\bomega\right)%
    &= \frac{1}{2}\log\frac{\tau_1^2}{(\tau_{1}^*)^2} - \frac{1}{2}\left\{1-\frac{(\tau_{1}^*)^2}{\tau_{1}^2}\right\} + \frac{1}{2\tau_1^2}\int_\calD \left\{\mu_{1}^*(\bs)-\mu_{1}(\bs)\right\}^2 p_1\left(\bs|\bomega^*\right)d\bs\tag{S1}\label{pt1}\\
    \hspace{15pt}&-\int_{\calD}\left\{\bX(\bs)^{\top}\left(\bdelta_1-\bdelta_{1}^*\right) + V_{1}(\bs)-V_{1}^*(\bs)+\phi_{1}U_{1}(\bs)-\phi_{1}^*U_{1}^*(\bs)\right\}p_1(\bs|\bomega)d\bs \tag{S2}\label{pt2}\\
    \hspace{15pt}&+\log\left[\frac{\int_\calD \exp \left\{\bX(\bs)^{\top}\bdelta_1 + V_1(\bs)+\phi_1U_1(\bs)\right\}d\bs}{\int_\calD \exp \left\{\bX(\bs)^{\top}\bdelta_{1}^* + V_{1}^*(\bs)+\phi_{1}^*U_{1}^*(\bs)\right\}d\bs} \right]\tag{S3}\label{pt3} \\
    \hspace{15pt}&+\log\frac{\mathbb{P}(A=1|\bomega^*)}{\mathbb{P}(A=1|\bomega)}. \tag{S4}\label{pt4}
\end{align}

Let the $\ell_\infty$-norm for a bounded function $h\in\calC(\calD)$ as $\|h\|_\infty = \sup_{\bs\in\calD}|h(\bs)|$ and denote the $\ell_2$ norm of a vector $\bv$ as $\|\bv\|_2$. 
For \eqref{pt1}, we clearly have
\begin{multline}
    \frac{1}{2}\log\frac{\tau_1^2}{(\tau_{1}^*)^2} - \frac{1}{2}\left(1-\frac{(\tau_{1}^*)^2}{\tau_{1}^2}\right) + \frac{1}{2\tau_1^2}\int_\calD \left\{\mu_{1}^*(\bs)-\mu_{1}(\bs)\right\}^2 p_1\left(\bs|\bomega^*\right)d\bs\\
    \leq \frac{1}{2}\log\frac{\tau_1^2}{(\tau_{1}^*)^2} - \frac{1}{2}\left(1-\frac{(\tau_{1}^*)^2}{\tau_{1}^2}\right) + \frac{\|\mu_{1}^*-\mu_{1}\|_\infty^2}{2\tau_1^2}. \tag{S5}\label{rhs}
\end{multline}
By using the equivalence of a norm \citep{gockenbach2011finite}, there exists $\Tilde{C}_1 >0$ such that $\|\bbeta_{1}^*-\bbeta_1\|_\infty \leq \Tilde{C}_1 \|\bbeta_{1}^* - \bbeta_1\|_2$. Using the assumption that $\|\bX(\bs)\|\leq C$ for all $\bs\in\calD$, Cauchy-Schwarz inequality, and triangle inequality, we have the following.
\begin{align*}
    \resizebox{0.98\hsize}{!}{$\|\mu_{1}^* - \mu_1\|_\infty = \|\alpha_1-\alpha_{1}^* + \bX^{\top}\left(\bbeta_1-\bbeta_{1}^*\right) + U_1-U_{1}^* \|_\infty \leq |\alpha_1-\alpha_{1}^*| + C_1\|\bbeta_{1}-\bbeta_{1}^*\|_2 + \|U_1 - U_{1}^*\|_\infty$}
\end{align*}
where $C_1=C\cdot\Tilde{C}_1$. By the LMC and triangle inequality, 
$$\|U_1 - U_{1}^*\|_\infty \leq \|\Tilde{U}_1-\Tilde{U}_1^*\|_\infty + \|\gamma_u\Tilde{U}_0-\gamma_u^*\Tilde{U}_0^*\|_\infty.$$
Therefore, since all the terms in the right hand side of \eqref{rhs} are bounded, for all $\varepsilon>0$, there exists $\nu_1>0$ such that $\bomega \in B_{\nu_1}$ implies
\begin{equation*}
    \frac{1}{2}\log\frac{\tau_1^2}{(\tau_{1}^*)^2} - \frac{1}{2}\left(1-\frac{(\tau_{1}^*)^2}{\tau_{1}^2}\right) + \frac{1}{2\tau_1^2}\int_\calD \left\{\mu_{1}^*(\bs)-\mu_{1}(\bs)\right\}^2 p_1\left(\bs|\bomega^*\right)d\bs < \varepsilon/8.
\end{equation*}

Second, let us proceed to \eqref{pt2}. Due to the norm equivalence, Cauchy-Schwarz inequality, and triangle inequality, we have
\begin{align*}
    &\bX(\bs)^{\top}\left(\bdelta_1-\bdelta_{1}^*\right) + V_{1}(\bs)-V_{1}^*(\bs)+\phi_{1}U_{1}(\bs)-\phi_{1}^*U_{1}^*(\bs) \\
    &\leq \left|\bX(\bs)^{\top}\left(\bdelta_1-\bdelta_{1}^*\right) + V_{1}(\bs)-V_{1}^*(\bs)+\phi_{1}U_{1}(\bs)-\phi_{1}^*U_{1}^*(\bs)\right| \\
    &\leq \Tilde{C}_2\|\bdelta_1-\bdelta_{1}^*\|_2 + \|V_1 - V_{1}^*\|_\infty + \|\phi_{1}U_{1}-\phi_{1}^*U_{1}^*\|_\infty.
\end{align*}
By the LMC, we have $\|V_1 - V_{1,*}\|_\infty \leq \|\Tilde{V}_1-\Tilde{V}_1^*\|_\infty + \|\gamma_v\Tilde{V}_0-\gamma_v^*\Tilde{V}_0^*\|_\infty$. By the LMC and triangle inequality,
$$|\phi_{1}U_{1}-\phi_{1,*}U_{1,*}\|_\infty \leq \|\phi_1\Tilde{U}_1 - \phi_1^*\Tilde{U}_1^*\|_\infty + \|\phi_1\gamma_u\Tilde{U}_0 - \phi_1^*\gamma_u^*\Tilde{U}_0^*\|_\infty.$$
Then, there exists $\nu_2>0$ such that for $\bomega \in B_{\nu_2}$,
\begin{equation*}
    \int_\calD \left\{\bX(\bs)^{\top}\left(\bdelta_1-\bdelta_{1}^*\right) + V_{1}(\bs)-V_{1}^*(\bs)+\phi_{1}U_{1}(\bs)-\phi_{1}^*U_{1}^*(\bs)\right\}p_1\left(\bs|\bomega\right)d\bs < \varepsilon / 8.
\end{equation*}

For \eqref{pt3}, let $$\Lambda_a(\bomega)=\log\left[\int_{\calD}\exp\left\{\bX(\bs)^{\top}\bdelta_a + V_a(\bs) + \phi_aU_a(\bs)\right\}d\bs\right].$$
Since $\exp\left\{\bX(\bs)^{\top}\bdelta_a + V_a(\bs) + \phi_aU_a(\bs)\right\}$ is a continuous function in $\bomega$, so is $\Lambda_a$. By using this continuity, we can say that for all $\varepsilon>0$, there exists $\nu_3 > 0$ such that $\bomega\in B_{\nu_3}$ implies
\begin{equation*}
    \log\left[\frac{\int_\calD \exp \left\{\bX(\bs)^{\top}\bdelta_1 + V_1(\bs)+\phi_{1}U_{1}(\bs)\right\}d\bs}{\int_\calD \exp \left\{\bX(\bs)^{\top}\bdelta_{1}^* + V_{1}^*(\bs)+\phi_{1}^*U_{1}^*(\bs)\right\}d\bs}\right] < \varepsilon/8.
\end{equation*}
In a similar sense, for \eqref{pt4}, since $\lambda_0(\bs)$ and $\lambda_1(\bs)$ are continuous functions in $\bomega$ and so is $\mathbb{P}(A=1|\bomega)$. Therefore, for all $\varepsilon>0$, there exists $\nu_4 > 0$ such that $\bomega\in B_{\nu_4}$ implies $$\log\frac{\mathbb{P}(A=1|\bomega^*)}{\mathbb{P}(A=1|\bomega)}<\varepsilon/8.$$

Finally, by taking $\nu = \min\left\{\nu_1, \nu_2, \nu_3, \nu_4\right\}$, $\bomega \in B_\nu$ implies
\begin{equation*}
    K_1\left(\bomega^*, \bomega\right) < \varepsilon/2.
\end{equation*}
It can be verified that this result holds for $K_0(\bomega^*, \bomega)$ by following similar steps and we have $K\left(\bomega^*, \bomega\right) < \varepsilon$. 

To complete the proof, we show that the prior probability on $\left\{\bomega: \bomega\in B_{\nu}(\bomega^*)\right\}$ is positive. If we assign priors to ${\btheta}_\calM$ as described in Assumption 4, the prior distribution of $\btheta_\calM$ has a positive probability for $\bomega\in B_\nu(\bomega^*)$. For the prior positivity of $\calW$, we require the following lemma.
\begin{lemma}
    Let $U(\bs)$ be a mean-zero Gaussian process on $\calD\subseteq\mathbb{R}^2$ with a Matérn covariance kernel parametrized by variance $\sigma^2$, spatial dependence $\rho$, and smoothness parameter $\kappa$. Then, $U(\bs)$ has a continuous sample path if $\kappa\geq1/2$.
    \label{kolmogorov}
    \begin{proof}
    The proof uses a result from \cite{adler2009random} that requires a bound on $$\mbox{E}\left\{|U(\bs)-U(\bs')|^2\right\}=2\sigma^2\left\{1-\calR(h;\rho,\kappa)\right\},$$ where $h=\|\bs-\bs'\|$ and $\calR(h;\rho,\kappa)$ has the form (10). Since $\frac{\partial h^{\kappa}\calK_\kappa(h)}{\partial h} = -h^{\kappa}\calK_{\kappa-1}(h)$ (\citealp[10.29.2 of ][]{olver2010nist}),
    $$\bigg|\frac{\partial\calR(h;\rho,\kappa)}{\partial h}\bigg| = M_1\cdot \left(\frac{h}{\rho}\right)^{\kappa}\calK_{\kappa-1}\left(\frac{h}{\rho}\right)$$
    for some constant $M_1>0$. Since $\calK_{\kappa-1}(h) < \calK_{\kappa}(h)$ when $\kappa>1/2$ (\citealp[3.3 of ][]{laforgia1991bounds}),
    $$\bigg|\frac{\partial\calR(h;\rho,\kappa)}{\partial h}\bigg| < M_1\cdot\left(\frac{h}{\rho}\right)^{\kappa}\calK_{\kappa}\left(\frac{h}{\rho}\right) = M_2\cdot\calR(h;\rho,\kappa)\leq M_2$$
    for some constant $M_2>0$. When $\kappa=1/2$, $\frac{\partial\calR(h;\rho,\kappa)}{\partial h} \propto \exp(-h/\rho)$, which is bounded for all $h>0$. Therefore, $\frac{\partial \calR(h;\rho,\kappa)}{\partial h}$ is finite for $h>0$ when $\kappa\geq1/2$.

    Let $C_{\text{sup}}=\sup_{h>0,\kappa\geq1/2}\left|\frac{\partial\calR(h;\rho,\kappa)}{\partial h}\right|$. By the mean value theorem and the boundedness of the derivative of $\calR(h;\rho,\kappa)$, we have
    \begin{equation*}
        \mbox{E}\left\{|U(\bs)-U(\bs')|^2\right\} = -\frac{\partial \calR(h;\rho,\kappa)}{\partial h}\bigg|_{h={h}_0}\cdot2\sigma^2 h \leq 2\sigma^2C_{\text{sup}}\cdot h, 
    \end{equation*}
    where $h_0$ lies in between 0 and $h$. Since $h\leq{|\log(h)|^{-2}}$ for $h<1$,
    \begin{equation*}
        \mbox{E}\left\{|U(\bs)-U(\bs')|^2\right\} \leq \frac{2\sigma^2C_{\text{sup}}}{|\log(h)|^2} 
    \end{equation*}
    and by Theorem 1.4.1 from \cite{adler2009random}, $U(\bs)$ has a continuous sample path.
\end{proof}
\end{lemma}

We establish prior positivity for $\calW$ using Lemma~\ref{kolmogorov} by conditioning on the events $\{\kappa_u \geq 1/2\}$ and $\{\kappa_v \geq 1/2\}$ following Theorem 4 of \cite{ghosal2006posterior}, which assumes the continuity of the sample paths of Gaussian processes. However, by Assumption 4 that we assign the smoothness parameters with the priors having positive mass on those events, the prior positivity of $\calW$ holds marginally over $\btheta_\calC$ and subsequently
\begin{equation*}
    \Pi\left\{\bomega : \bomega\in B_\nu (\bomega^*) \right\} > 0.
\end{equation*}
Since $\bomega\in B_\nu$ implies $K\left(\bomega^*, \bomega\right) < \varepsilon$ for all $\varepsilon>0$, we have
\begin{equation*}
    \Pi\left\{\bomega : K\left(\bomega^*, \bomega\right) < \varepsilon \right\} > 0.
\end{equation*}
Therefore, by \cite{schwartz1965bayes}, the posterior is weakly consistent and this completes the proof of Theorem 2.

\subsection{Derivations of the Gibbs sampler}\label{supp:gibbs}

Let us denote $q(\cdot|\cdot)$ a full conditional density, $\calL(\cdot|\cdot)$ a likelihood, and $\pi(\cdot)$ a marginal prior density. Let us denote $\cdot|\mbox{rest}$ a full conditional distribution of a parameter conditioned on the remaining parameters. 

\vspace{10pt}
\noindent
{\bf Full conditional distribution for $\bbeta_0,\bbeta_1$}

\noindent
Let the prior be $\bbeta_a\sim\text{MVN}\left(\mathbf{0}_p,c_{\beta}^2\bI_p\right)$ for $a\in\left\{0,1\right\}$. Let $\bY_a$ be a $n\times1$ vector of potential outcomes $Y_a(\bs_i)$ for $i=1,\cdots,n$ and $\bX_\calS$ be a $n\times p$ covariate matrix with $i^{\text{th}}$ row corresponding to $\bX(\bs_i)$, respectively. Note that $\bY_a$ has complete entries since the missing values are imputed at every step of MCMC iterations. Let $\bH$ be a $n\times G$ matrix where $(i,g)$ entry corresponds to 1 if $i^{\text{th}}$ observation belongs to $\calD_g$ for some $g=1,\cdots,G$, and 0 otherwise. Then,
\small{\begin{align*}
    &q\left(\bbeta_a|\mbox{rest}\right)\\
    &\propto \calL\left(\bY_a|\bbeta_a,\mbox{rest}\right)\pi\left(\bbeta_a\right)\\
    &\propto \exp\left\{-\frac{\left(\bY_a-\bH\bU_a-\alpha_a\mathbf{1}_n-\bX_\calS\bbeta_a\right)^{\top}\left(\bY_a-\bH\bU_a-\alpha_a\mathbf{1}_n-\bX_\calS\bbeta_a\right)}{2\tau_a^2}\right\}\times\\
    &\hspace{15pt}\exp\left(-\frac{\bbeta_a^{\top}\bbeta_a}{2c_\beta^2}\right)\nonumber\\
    &\propto \exp\left\{-\frac{\bbeta_a^{\top}\bX_\calS^{\top}\bX_\calS\bbeta_a - 2\bbeta_a^{\top}\bX_\calS^{\top}\left(\bY_a - \bH\bU_a-\alpha_a\mathbf{1}_n\right)}{2\tau_a^2}\right\}\exp\left(-\frac{\bbeta_a^{\top}\bbeta_a}{2c_\beta^2}\right)\\
    &\propto \exp\left[-\frac{1}{2}\left\{\bbeta_a^{\top}\left(\frac{\bX_\calS^{\top}\bX_\calS}{\tau_a^2} + \frac{\mathbb{I}_p}{c_\beta^2}\right)\bbeta_a-\frac{2\bbeta_a^{\top}\bX_\calS^{\top}\left(\bY_a-\bH\bU_a-\alpha_a\mathbf{1}_n\right)}{\tau_a^2}\right\}\right].
\end{align*}}
Thus, 
\begin{equation*}
\bbeta_a|\text{rest}\sim\mbox{MVN}\left(\bC_a^{-1}\bb_a,\bC_a^{-1}\right)
\end{equation*}
where $\bb_a=\bX_\calS^{\top}\left(\bY_a-\bH\bU_a-\alpha_a\mathbf{1}_n\right)/\tau_a^2$ and $\bC_a={\bX_\calS^{\top}\bX_\calS}/{\tau_a^2} + {\mathbb{I}_p}/{c_\beta^2}$. 

\vspace{10pt}
\noindent
{\bf Full conditional distribution for $\alpha_0,\alpha_1$}

\noindent
Let the prior be $\alpha_a\sim\text{Normal}\left(0,c_{\alpha}^2\right)$ for $a\in\left\{0,1\right\}$. Then,
\small{\begin{align*}
    &q\left(\alpha_a|\mbox{rest}\right)\\
    &\propto \calL\left(\bY_a|\alpha_a,\mbox{rest}\right)\pi\left(\alpha_a\right)\\
    &\propto \exp\left\{-\frac{\left(\bY_a-\bH\bU_a-\alpha_a\mathbf{1}_n-\bX_\calS\bbeta_a\right)^{\top}\left(\bY_a-\bH\bU_a-\alpha_a\mathbf{1}_n-\bX_\calS\bbeta_a\right)}{2\tau_a^2}\right\}\exp\left(-\frac{\alpha_a^2}{2c_\alpha^2}\right)\\
    &\propto \exp\left\{-\frac{n\cdot\alpha_a^2 - 2\alpha_a\cdot\mathbf{1}_n^{\top}\left(\bY_a - \bH\bU_a-\bX_\calS\bbeta_a\right)}{2\tau_a^2}\right\}\exp\left(-\frac{\alpha_a^2}{2c_\alpha^2}\right)\\
    &\propto \exp\left[-\frac{1}{2}\left\{\left(\frac{n}{\tau_a^2} + \frac{1}{c_\alpha^2}\right)\alpha_a^2-\frac{2\cdot\mathbf{1}_n^{\top}\left(\bY_a - \bH\bU_a-\bX_\calS\bbeta_a\right)}{\tau_a^2}\alpha_a\right\}\right].
\end{align*}}
Thus, 
\begin{equation*}
\alpha_a|\text{rest}\sim\mbox{Normal}\left(b_a/C_a,1/C_a\right)
\end{equation*}
where $b_a=\mathbf{1}_n^{\top}\left(\bY_a - \bH\bU_a-\bX_\calS\bbeta_a\right)/\tau_a^2$ and $C_a=\left(\frac{n}{\tau_a^2} + \frac{1}{c_\alpha^2}\right)$.

\vspace{10pt}
\noindent
{\bf Full conditional distribution for $\bdelta_0,\bdelta_1$}

\noindent
Let the prior be $\bdelta_a\sim\text{MVN}\left(\mathbf{0}_p,c_{\delta}^2\bI_p)\right)$. Let $\bL_a$ be a $G\times1$ vector with its $g^{\text{th}}$ entry being $\log(\lambda_{a,g})$ and $\bX_\calG$ be a $G\times p$ matrix with every $g^{\text{th}}$ row corresponding to $\bX_g$. Then,
\small{\begin{align*}
    &q\left(\bdelta_a|\mbox{rest}\right)\\
    &\propto \calL\left(\bL_a|\bdelta_a,\mbox{rest}\right)\pi\left(\bdelta_a\right)\\
    &\propto \exp\left\{-\frac{\left(\bL_a - \eta_a\mathbf{1}_G - \bX_\calG\bdelta_a - \bV_a - \phi_a\bU_a\right)^{\top}\left(\bL_a - \eta_a\mathbf{1}_G - \bX_\calG\bdelta_a - \bV_a - \phi_a\bU_a\right)}{2\tau_{\psi_a}^2}\right\}\times \\
    &\hspace{15pt}\exp\left(-\frac{\bdelta_a^2}{2c_\delta^2}\right)\nonumber\\
    &\propto \exp\left\{-\frac{\bdelta_a^{\top}\bX_\calG^{\top}\bX_\calG\bdelta_a - 2\bdelta_a^{\top}\bX_\calG^{\top}\left(\bL_a - \eta_a\mathbf{1}_G -\bV_a - \phi_a\bU_a\right)}{2\tau_{\psi_a}^2}\right\}\exp\left(-\frac{\bdelta_a^{\top}\bdelta_a}{2c_\delta^2}\right)\\
    &\propto \exp\left[-\frac{1}{2}\left\{\bdelta_a^{\top}\left(\frac{\bX_\calG^{\top}\bX_\calG}{\tau_{\psi_a}^2}+\frac{\mathbb{I}_p}{c_\delta^2}\right)\bdelta_a-\frac{2\bdelta_a^{\top}\bX_\calG^{\top}\left(\bL_a-\eta_a\mathbf{1}_G-\bV_a-\phi_a\bU_a\right)}{\tau_{\psi_a}^2}\right\}\right].
\end{align*}}
Thus, 
\begin{equation*}
\bdelta_a|\text{rest}\sim\mbox{MVN}\left(\bC_a^{-1}\bb_a,\bC_a^{-1}\right)
\end{equation*}
where $\bb_a=\bX_\calG^{\top}\left(\bL_a-\eta_a\mathbf{1}_G-\bV_a-\phi_a\bU_a\right)/\tau_{\psi_a}^2$ and $\bC_a={\bX_\calG^{\top}\bX_\calG}/{\tau_{\psi_a}^2} + {\mathbb{I}_p}/{c_\delta^2}$.

\vspace{10pt}
\noindent
{\bf Full conditional distribution for $\eta_0,\eta_1$}

\noindent
Let the prior be $\eta_a\sim\text{Normal}\left(0,c_{\eta}^2\right)$. Then,
\small{\begin{align*}
    &q\left(\eta_a|\mbox{rest}\right)\\
    &\propto \calL\left(\bL_a|\eta_a,\mbox{rest}\right)\pi\left(\eta_a\right)\\
    &\propto \exp\left\{-\frac{\left(\bL_a - \eta_a\mathbf{1}_G - \bX_\calG\bdelta_a - \bV_a - \phi_a\bU_a\right)^{\top}\left(\bL_a - \eta_a\mathbf{1}_G - \bX_\calG\bdelta_a - \bV_a - \phi_a\bU_a\right)}{2\tau_{\psi_a}^2}\right\}\times \\
    &\hspace{15pt}\exp\left(-\frac{\eta_a^2}{2c_\eta^2}\right)\nonumber\\
    &\propto \exp\left\{-\frac{G\cdot\eta_a^2-2\cdot\mathbf{1}_G^{\top}\left(\bL_a-\bX_\calG\bdelta_a-\bV_a-\phi_a\bU_a\right)}{2\tau_{\psi_a}^2}\right\}\exp\left(-\frac{\eta_a^2}{2c_\eta^2}\right)\\
    &\propto \exp\left[-\frac{1}{2}\left\{\left(\frac{G}{\tau_{\psi_a}^2}+\frac{1}{c_\eta^2}\right)\eta_a^2-\frac{2\cdot\mathbf{1}_G^{\top}\left(\bL_a-\bX_\calG\bdelta_a-\bV_a-\phi_a\bU_a\right)}{\tau_{\psi_a}^2}\eta_a\right\}\right].
\end{align*}}
Thus, 
\begin{equation*}
\eta_a|\text{rest}\sim\mbox{Normal}\left(b_a/C_a,1/C_a\right)
\end{equation*}
where $b_a=\mathbf{1}_G^{\top}\left(\bL_a-\bX_\calG\bdelta_a-\bV_a-\phi_a\bU_a\right)/\tau_{\psi_a}^2$ and $C_a=\left(\frac{G}{\tau_{\psi_a}^2} + \frac{1}{c_\eta^2}\right)$.

\vspace{10pt}
\noindent
{\bf Full conditional distribution for $\phi_0,\phi_1$}

\noindent
Let the prior be $\phi_a\sim\text{Normal}\left(0,c_{\phi}^2\right)$. Then,
\small{\begin{align*}
    &q\left(\phi_a|\mbox{rest}\right)\\
    &\propto \calL\left(\bL_a|\phi_a,\mbox{rest}\right)\pi\left(\phi_a\right)\\
    &\propto \exp\left\{-\frac{\left(\bL_a - \eta_a\mathbf{1}_G - \bX_\calG\bdelta_a - \bV_a - \phi_a\bU_a\right)^{\top}\left(\bL_a - \eta_a\mathbf{1}_G - \bX_\calG\bdelta_a - \bV_a - \phi_a\bU_a\right)}{2\tau_{\psi_a}^2}\right\}\times \\
    &\hspace{15pt}\exp\left(-\frac{\phi_a^2}{2c_\phi^2}\right)\nonumber\\
    &\propto \exp\left\{-\frac{\bU_a^{\top}\bU_a\phi_a^2-2\phi_a\cdot\bU_a^{\top}\left(\bL_a-\eta\mathbf{1}_G-\bX_\calG\bdelta_a-\bV_a\right)}{2\tau_{\psi_a}^2}\right\}\exp\left(-\frac{\phi_a^2}{2c_\phi^2}\right)\\
    &\propto \exp\left[-\frac{1}{2}\left\{\left(\frac{\bU_a^{\top}\bU_a}{\tau_{\psi_a}^2}+\frac{1}{c_\phi^2}\right)\phi_a^2-\frac{2\cdot\bU_a^{\top}\left(\bL_a-\eta\mathbf{1}_G-\bX_\calG\bdelta_a-\bV_a\right)}{\tau_{\psi_a}^2}\phi_a\right\}\right].
\end{align*}}
Thus, 
\begin{equation*}
\phi_a|\text{rest}\sim\mbox{Normal}\left(b_a/C_a,1/C_a\right)
\end{equation*}
where $b_a=\bU_a^{\top}\left(\bL_a-\eta_a\mathbf{1}_G-\bX_\calG\bdelta_a-\bV_a\right)/\tau_{\psi_a}^2$ and $C_a=\left(\frac{\bU_a^{\top}\bU_a}{\tau_{\psi_a}^2} + \frac{1}{c_\eta^2}\right)$.

\vspace{10pt}
\noindent
{\bf Full conditional distribution for $\gamma_u$}

\noindent
Let the prior be $\gamma_u\sim\text{Normal}\left(0,c_\gamma^2\right)$. By LMC, $\bU_0 = \Tilde{\bU}_0$ and $\bU_1 = \Tilde{\bU}_1 + \gamma_u\Tilde{\bU}_0$. Let $\bmm_{\beta_1}=\alpha_1\mathbf{1}_n+\bX_\calS\bbeta_1$ and $\bmm_{\delta_1}=\eta_1\mathbf{1}_G+\bX_\calG\bdelta_1$ for brevity of notation. Then,
\small{\begin{align*}
    &q\left(\gamma_u|\mbox{rest}\right)\\
    &\propto \calL\left(\bY_1|\gamma_u,\mbox{rest}\right)f\left(\bL_1|\gamma_u,\mbox{rest}\right)\pi\left(\gamma_u\right)\\
    &\propto \exp\left\{-\frac{\left(\bY_1-\bmm_{\beta_1}-\bH\Tilde{\bU}_1-\gamma_u\cdot\bH\Tilde{\bU}_0\right)^{\top}\left(\bY_1-\bmm_{\beta_1}-\bH\Tilde{\bU}_1-\gamma_u\cdot\bH\Tilde{\bU}_0\right)}{2\tau_1^2}\right\}\times \\
    &\hspace{15pt} \exp\left\{-\frac{\left(\bL_1-\bmm_{\delta_1}-\bV_1-\phi_1\Tilde{\bU}_1-\gamma_u\cdot\phi_1\Tilde{\bU}_0\right)^{\top}\left(\bL_1-\bmm_{\delta_1}-\bV_1-\phi_1\Tilde{\bU}_1-\gamma_u\cdot\phi_1\Tilde{\bU}_0\right)}{2\tau_{\psi_1}^2}\right\}\times\nonumber \\
    &\hspace{15pt} \exp\left(-\frac{\gamma_u^2}{2c_\gamma^2}\right)\nonumber \\
    &\propto \exp\left\{-\frac{\Tilde{\bU}_0^{\top}\bH^{\top}\bH\Tilde{\bU}_0\cdot\gamma_u^2-2\gamma_u\cdot\Tilde{\bU}_0^{\top}\bH^{\top}\left(\bY_1-\bmm_{\beta_1}-\bH\Tilde{\bU}_1\right)}{2\tau_1^2}\right\}\times \\
    &\hspace{15pt}\exp\left\{-\frac{\phi_1^2\Tilde{\bU}_0^{\top}\Tilde{\bU}_0\cdot\gamma_u^2-2\gamma_u\cdot\phi_1\Tilde{\bU}_0^{\top}\left(\bL_1-\bmm_{\delta_1}-\bV_1-\phi_1\Tilde{\bU}_1\right)}{2\tau_{\psi_1}^2}\right\}\times\exp\left(-\frac{\gamma_u^2}{2c_\gamma^2}\right)\nonumber \\
    &\propto \exp\left\{-\frac{1}{2}\left(\frac{\Tilde{\bU}_0^{\top}\bH^{\top}\bH\Tilde{\bU}_0}{\tau_1^2}+\frac{\phi_1^2\Tilde{\bU}_0^{\top}\Tilde{\bU}_0}{\tau_{\psi_1}^2}+\frac{1}{c_\gamma^2}\right)\gamma_u^2\right\} \times \\
    &\hspace{15pt}\exp\left[-\left\{\frac{\Tilde{\bU}_0^{\top}\bH^{\top}\left(\bY_1-\bmm_{\beta_1}-\bH\Tilde{\bU}_1\right)}{\tau_1^2} + \frac{\phi_1\Tilde{\bU}_0^{\top}\left(\bL_1-\bmm_{\delta_1}-\bV_1-\phi_1\Tilde{\bU}_1\right)}{\tau_{\psi_1}^2}\right\}\gamma_u\right]. \nonumber
\end{align*}}
Thus, 
\begin{equation*}
\gamma_u|\text{rest}\sim\mbox{Normal}\left(b/C,1/C\right)
\end{equation*}
where $$b={\Tilde{\bU}_0^{\top}\bH^{\top}\left(\bY_1-\bmm_{\beta_1}-\bH\Tilde{\bU}_1\right)}/{\tau_1^2} + {\phi_1\Tilde{\bU}_0^{\top}\left(\bL_1-\bmm_{\delta_1}-\bV_1-\phi_1\Tilde{\bU}_1\right)}/{\tau_{\psi_1}^2}$$ and $$C=\left(\frac{\Tilde{\bU}_0^{\top}\bH^{\top}\bH\Tilde{\bU}_0}{\tau_1^2}+\frac{\phi_1^2\Tilde{\bU}_0^{\top}\Tilde{\bU}_0}{\tau_{\psi_1}^2}+\frac{1}{c_\gamma^2}\right).$$

\vspace{10pt}
\noindent
{\bf Full conditional distribution for $\gamma_v$}

\noindent
Let the prior be $\gamma_v\sim\text{Normal}\left(0,c_\gamma^2\right)$. By LMC, $\bV_0 = \Tilde{\bV}_0$ and $\bV_1 = \Tilde{\bV}_1 + \gamma_v\Tilde{\bV}_0$. Then,
\small{\begin{align*}
    &q\left(\gamma_v|\mbox{rest}\right)\\
    &\propto \calL\left(\bL_1|\gamma_v,\mbox{rest}\right)\pi\left(\gamma_v\right)\\
    &\propto \exp\left\{-\frac{\left(\bL_1-\bmm_{\delta_1}-\Tilde{\bV}_1-\gamma_v\Tilde{\bV}_0-\phi_1\bU_1\right)^{\top}\left(\bL_1-\bmm_{\delta_1}-\Tilde{\bV}_1-\gamma_v\Tilde{\bV}_0-\phi_1\bU_1\right)}{2\tau_{\psi_1}^2}\right\}\times\nonumber \\
    &\hspace{15pt} \exp\left(-\frac{\gamma_v^2}{2c_\gamma^2}\right)\nonumber \\
    &\propto\exp\left\{-\frac{\Tilde{\bV}_0^{\top}\Tilde{\bV}_0\cdot\gamma_v^2-2\gamma_v\cdot\Tilde{\bV}_0^{\top}\left(\bL_1-\bmm_{\delta_1}-\Tilde{\bV}_1-\phi_1{\bU}_1\right)}{2\tau_{\psi_1}^2}\right\}\times\exp\left(-\frac{\gamma_v^2}{2c_\gamma^2}\right) \\
    &\propto\exp\left[-\frac{1}{2}\left\{\left(\frac{\Tilde{\bV}_0^{\top}\Tilde{\bV}_0}{\tau_{\psi_1}^2}+\frac{1}{c_{\gamma}^2}\right)\gamma_v^2 + \frac{2\Tilde{\bV}_0^{\top}\left(\bL_1-\bmm_{\delta_1}-\Tilde{\bV}_1-\phi_1{\bU}_1\right)}{\tau_{\psi_1}^2}\gamma_u\right\}\right].
\end{align*}}
Thus, 
\begin{equation*}
\gamma_u|\text{rest}\sim\mbox{Normal}\left(b/C,1/C\right)
\end{equation*}
where $b=\Tilde{\bV}_0^{\top}\left(\bL_1-\bmm_{\delta_1}-\Tilde{\bV}_1-\phi_1{\bU}_1\right)/{\tau_{\psi_1}^2}$ and $C=\left(\frac{\Tilde{\bV}_0^{\top}\Tilde{\bV}_0}{\tau_{\psi_1}^2}+\frac{1}{c_{\gamma}^2}\right)$.

\vspace{10pt}
\noindent
{\bf Full conditional distribution for $\Tilde{\bU}_0$}

\noindent
Note that $\Tilde{\bU}_0\sim\text{MVN}\left\{\mathbf{0}_G, \sigma_{u,0}^2\bR(\rho_u,\kappa_u)\right\}$. Let $\bmm_{\beta_0}=\alpha_0\mathbf{1}_n+\bX_\calS\bbeta_0$ and $\bmm_{\delta_0}=\eta_0\mathbf{1}_G+\bX_\calG\bdelta_0$ for brevity. Then,
\small{\begin{align*}
    &q\left(\Tilde{\bU}_0|\text{rest}\right) \\
    &\propto \calL\left(\bY_0|\Tilde{\bU}_0, \text{rest}\right)\calL\left(\bY_1|\Tilde{\bU}_0, \text{rest}\right)\calL\left(\bL_0|\Tilde{\bU}_0, \text{rest}\right)\calL\left(\bL_1|\Tilde{\bU}_0, \text{rest}\right)\pi\left(\Tilde{\bU}_0\right) \\
    &\propto \exp\left\{-\frac{\left(\bY_0-\bmm_{\beta_0}-\bH\Tilde{\bU}_0\right)^{\top}\left(\bY_0-\bmm_{\beta_0}-\bH\Tilde{\bU}_0\right)}{2\tau_0^2}\right\}\times \\
    &\hspace{15pt}\exp\left\{-\frac{\left(\bY_1-\bmm_{\beta_1}-\bH\Tilde{\bU}_1-\gamma_u\bH\Tilde{\bU}_0\right)^{\top}\left(\bY_0-\bmm_{\beta_1}-\bH\Tilde{\bU}_1-\gamma_u\bH\Tilde{\bU}_0\right)}{2\tau_1^2}\right\}\times \nonumber \\
    &\hspace{15pt}\exp\left\{-\frac{\left(\bL_0-\bmm_{\delta_0}-\bV_0-\phi_0\Tilde{\bU}_0\right)^{\top}\left(\bL_0-\bmm_{\delta_0}-\bV_0-\phi_0\Tilde{\bU}_0\right)}{2\tau_{\psi_0}^2}\right\} \times \nonumber \\
    &\hspace{15pt}\exp\left\{-\frac{\left(\bL_1-\bmm_{\delta_1}-\bV_1-\phi_1\Tilde{\bU}_1-\phi_1\gamma_u\Tilde{\bU}_0\right)^{\top}\left(\bL_1-\bmm_{\delta_1}-\bV_1-\phi_1\Tilde{\bU}_1-\phi_1\gamma_u\Tilde{\bU}_0\right)}{2\tau_{\psi_1}^2}\right\}\times \nonumber \\
    &\hspace{15pt}\exp\left\{-\frac{\Tilde{\bU}_0^{\top}\bR^{-1}(\rho_u,\kappa_u)\Tilde{\bU}_0}{2\sigma_{u,0}^2}\right\} \nonumber \\
    &\propto \exp\left\{-\frac{\Tilde{\bU}_0^{\top}\bH^{\top}\bH\Tilde{\bU}_0-2\Tilde{\bU}_0^{\top}\bH^{\top}\left(\bY_0-\bmm_{\beta_0}\right)}{2\tau_0^2}\right\}\times \\
    &\hspace{15pt}\exp\left\{-\frac{\gamma_u^2\Tilde{\bU}_0\bH^{\top}\bH\Tilde{\bU}_0 -2\gamma_u\Tilde{\bU}_0^{\top}\bH^{\top}\left(\bY_1-\bmm_{\beta_1}-\bH\Tilde{\bU}_1\right)}{2\tau_1^2}\right\}\times \nonumber \\
    &\hspace{15pt}\exp\left\{-\frac{\phi_0^2\Tilde{\bU}_0^{\top}\Tilde{\bU}_0-2\phi_0\Tilde{\bU}_0^{\top}\left(\bL_0-\bmm_{\delta_0}-\bV_0\right)}{2\tau_{\psi_0}^2}\right\} \times \nonumber \\
    &\hspace{15pt}\exp\left\{-\frac{(\phi_1\gamma_u)^2\Tilde{\bU}_0^{\top}\Tilde{\bU}_0-2\phi_1\gamma_u\Tilde{\bU}_0^{\top}\left(\bL_1-\bmm_{\delta_1}-\bV_1-\phi_1\Tilde{\bU}_1\right)}{2\tau_{\psi_1}^2}\right\}\exp\left\{-\frac{\Tilde{\bU}_0^{\top}\bR^{-1}(\rho_u,\kappa_u)\Tilde{\bU}_0}{2\sigma_{u,0}^2}\right\} \nonumber \\
    &\propto \exp\left[-\frac{1}{2}\left\{\Tilde{\bU}_0^{\top}\left(\frac{\bH^{\top}\bH}{\tau_0^2}+\frac{\gamma_u^2\bH^{\top}\bH}{\tau_1^2}+\frac{\phi_0^2\mathbb{I}_G}{\tau_{\psi_0}^2}+\frac{\phi_1^2\gamma_u^2\mathbb{I}_G}{\tau_{\psi_1}^2}+\frac{\bR^{-1}(\rho_u,\kappa_u)}{\sigma_{u,0}^2}\right)\Tilde{\bU}_0\right\}\right]\times \\
    &\hspace{15pt}\exp\left[-\Tilde{\bU}_0^{\top}\left\{\frac{\bH^{\top}\left(\bY_0-\bmm_{\beta_0}\right)}{\tau_0^2}+\frac{\gamma_u\bH^{\top}\left(\bY_1-\bmm_{\beta_1}-\bH\Tilde{\bU}_1\right)}{\tau_1^2}+\frac{\phi_0\left(\bL_0-\bmm_{\delta_0}-\bV_0\right)}{\tau_{\psi_0}^2}\right\}\right]\times \nonumber \\
    &\hspace{15pt}\exp\left[-\Tilde{\bU}_0^{\top}\left\{\frac{\phi_1\gamma_u\left(\bL_1-\bmm_{\delta_1}-\bV_1-\phi_1\Tilde{\bU}_1\right)}{\tau_{\psi_1}^2}\right\}\right]. \nonumber
\end{align*}}
Thus, $\Tilde{\bU}_0|\text{rest}\sim\text{MVN}\left(\bC^{-1}\bb, \bC^{-1}\right)$ where 
\begin{equation*}
    \scalebox{1.05}{$\bb=\frac{\bH^{\top}\left(\bY_0-\bmm_{\beta_0}\right)}{\tau_0^2}+\frac{\gamma_u\bH^{\top}\left(\bY_1-\bmm_{\beta_1}-\bH\Tilde{\bU}_1\right)}{\tau_1^2}+\frac{\phi_0\left(\bL_0-\bmm_{\delta_0}-\bV_0\right)}{\tau_{\psi_0}^2}+\frac{\phi_1\gamma_u\left(\bL_1-\bmm_{\delta_1}-\bV_1-\phi_1\Tilde{\bU}_1\right)}{\tau_{\psi_1}^2}$}
\end{equation*}
and
$$\bC=\frac{\bH^{\top}\bH}{\tau_0^2}+\frac{\gamma_u^2\bH^{\top}\bH}{\tau_1^2}+\frac{\phi_0^2\mathbb{I}_G}{\tau_{\psi_0}^2}+\frac{\phi_1^2\gamma_u^2\mathbb{I}_G}{\tau_{\psi_1}^2}+\frac{\bR^{-1}(\rho_u,\kappa_u)}{\sigma_{u,0}^2}.$$

\vspace{10pt}
\noindent
{\bf Full conditional distribution for $\Tilde{\bU}_1$}

\noindent
Note that $\Tilde{\bU}_1\sim\text{MVN}\left\{\mathbf{0}_G, \sigma_{u,1}^2\bR(\rho_u,\kappa_u)\right\}$. Then,
\small{\begin{align*}
    &q\left(\Tilde{\bU}_0|\text{rest}\right) \\
    &\propto \calL\left(\bY_1|\Tilde{\bU}_1, \text{rest}\right)\calL\left(\bL_1|\Tilde{\bU}_1, \text{rest}\right)\pi\left(\Tilde{\bU}_1\right) \\
    &\propto\exp\left\{-\frac{\left(\bY_1-\bmm_{\beta_1}-\bH\Tilde{\bU}_1-\gamma_u\bH\Tilde{\bU}_0\right)^{\top}\left(\bY_0-\bmm_{\beta_1}-\bH\Tilde{\bU}_1-\gamma_u\bH\Tilde{\bU}_0\right)}{2\tau_1^2}\right\}\times \\
    &\hspace{15pt}\exp\left\{-\frac{\left(\bL_1-\bmm_{\delta_1}-\bV_1-\phi_1\Tilde{\bU}_1-\phi_1\gamma_u\Tilde{\bU}_0\right)^{\top}\left(\bL_1-\bmm_{\delta_1}-\bV_1-\phi_1\Tilde{\bU}_1-\phi_1\gamma_u\Tilde{\bU}_0\right)}{2\tau_{\psi_1}^2}\right\}\times \nonumber \\
    &\hspace{15pt}\exp\left\{-\frac{\Tilde{\bU}_1^{\top}\bR^{-1}(\rho_u,\kappa_u)\Tilde{\bU}_1}{2\sigma_{u,1}^2}\right\} \nonumber \\
    &\propto\exp\left\{-\frac{\Tilde{\bU}_1\bH^{\top}\bH\Tilde{\bU}_1 -2\Tilde{\bU}_1^{\top}\bH^{\top}\left(\bY_1-\bmm_{\beta_1}-\gamma_u\bH\Tilde{\bU}_0\right)}{2\tau_1^2}\right\}\times\\
    &\hspace{15pt}\exp\left\{-\frac{\phi_1^2\Tilde{\bU}_1^{\top}\Tilde{\bU}_1-2\phi_1\Tilde{\bU}_1^{\top}\left(\bL_1-\bmm_{\delta_1}-\bV_1-\phi_1\gamma_u\Tilde{\bU}_0\right)}{2\tau_{\psi_1}^2}\right\}\exp\left\{-\frac{\Tilde{\bU}_1^{\top}\bR^{-1}(\rho_u,\kappa_u)\Tilde{\bU}_1}{2\sigma_{u,1}^2}\right\} \nonumber \\
    &\propto \exp\left[-\frac{1}{2}\left\{\Tilde{\bU}_1^{\top}\left(\frac{\bH^{\top}\bH}{\tau_1^2}+\frac{\phi_1^2\mathbb{I}_G}{\tau_{\psi_1}^2}+\frac{\bR^{-1}(\rho_u,\kappa_u)}{\sigma_{u,1}^2}\right)\Tilde{\bU}_1\right\}\right]\times \\
    &\hspace{15pt}\exp\left[-\Tilde{\bU}_1^{\top}\left\{\frac{\bH^{\top}\left(\bY_1-\bmm_{\beta_1}-\gamma_u\bH\Tilde{\bU}_0\right)}{\tau_1^2}+\frac{\phi_1\left(\bL_1-\bmm_{\delta_1}-\bV_1-\phi_1\gamma_u\Tilde{\bU}_0\right)}{\tau_{\psi_1}^2}\right\}\right]. \nonumber 
\end{align*}}
Thus, $\Tilde{\bU}_1|\text{rest}\sim\text{MVN}\left(\bC^{-1}\bb, \bC^{-1}\right)$ where 
{\begin{equation*}
    \bb=\frac{\bH^{\top}\left(\bY_1-\bmm_{\beta_1}-\gamma_u\bH\Tilde{\bU}_0\right)}{\tau_1^2}+\frac{\phi_1\left(\bL_1-\bmm_{\delta_1}-\bV_1-\phi_1\gamma_u\Tilde{\bU}_0\right)}{\tau_{\psi_1}^2}
\end{equation*}}
and
$$\bC=\frac{\bH^{\top}\bH}{\tau_1^2}+\frac{\phi_1^2\mathbb{I}_G}{\tau_{\psi_1}^2}+\frac{\bR^{-1}(\rho_u,\kappa_u)}{\sigma_{u,1}^2}.$$

\vspace{10pt}
\noindent
{\bf Full conditional distribution for $\Tilde{\bV}_0$}

\noindent
Note that $\Tilde{\bV}_0\sim\text{MVN}\left\{\mathbf{0}_G, \sigma_{v,0}^2\bR(\rho_v,\kappa_v)\right\}$. Then,
\small{
\begin{align*}
    &q\left(\Tilde{\bV}_0|\text{rest}\right) \\
    &\propto \calL\left(\bL_0|\Tilde{\bV}_0, \text{rest}\right)\calL\left(\bL_1|\Tilde{\bV}_0, \text{rest}\right)\pi\left(\Tilde{\bV}_0\right) \\ 
    &\propto\exp\left\{-\frac{\left(\bL_0-\bmm_{\delta_0}-\Tilde{\bV}_0-\phi_0\bU_0\right)^{\top}\left(\bL_0-\bmm_{\delta_0}-\Tilde{\bV}_0-\phi_0\bU_0\right)}{2\tau_{\psi_0}^2}\right\} \times \\
    &\hspace{15pt}\exp\left\{-\frac{\left(\bL_1-\bmm_{\delta_1}-\Tilde{\bV}_1-\gamma_v\Tilde{\bV}_0-\phi_1\bU_1\right)^{\top}\left(\bL_1-\bmm_{\delta_1}-\Tilde{\bV}_1-\gamma_v\Tilde{\bV}_0-\phi_1\bU_1\right)}{2\tau_{\psi_1}^2}\right\} \times \nonumber \\
    &\hspace{15pt}\exp\left\{-\frac{\Tilde{\bV}_0^{\top}\bR^{-1}(\rho_v,\kappa_v)\Tilde{\bV}_0}{2\sigma_{v,0}^2}\right\} \nonumber \\
    &\propto \exp\left\{-\frac{\Tilde{\bV}_0^{\top}\Tilde{\bV}_0-2\Tilde{\bV}_0^{\top}\left(\bL_0-\bmm_{\delta_0}-\phi_0\bU_0\right)}{2\tau_{\psi_0}^2}\right\}\times \\
    &\hspace{15pt}\exp\left\{-\frac{\gamma_v^2\Tilde{\bV}_0^{\top}\Tilde{\bV}_0-2\gamma_v\Tilde{\bV}_0^{\top}\left(\bL_1-\bmm_{\delta_1}-\Tilde{\bV}_1-\phi_1\bU_1\right)}{2\tau_{\psi_1}^2}\right\}\exp\left\{-\frac{\Tilde{\bV}_0^{\top}\bR^{-1}(\rho_v,\kappa_v)\Tilde{\bV}_0}{2\sigma_{v,0}^2}\right\} \nonumber \\
    &\propto \exp\left\{-\frac{1}{2}\Tilde{\bV}_0^{\top}\left(\frac{\mathbb{I}_G}{\tau_{\psi_0}^2}+\frac{\gamma_v^2\mathbb{I}_G}{\tau_{\psi_1}^2}+\frac{\bR^{-1}(\rho_v,\kappa_v)}{\sigma_{v,0}^2}\right)\Tilde{\bV}_0\right\}\times \\
    &\hspace{15pt}\exp\left\{-\Tilde{\bV}_0^{\top}\left(\frac{\bL_0-\bmm_{\delta_0}-\phi_0\bU_0}{\tau_{\psi_0}^2}+\frac{\gamma_v\left(\bL_1-\bmm_{\delta_1}-\Tilde{\bV}_1-\phi_1\bU_1\right)}{\tau_{\psi_1}^2}\right)\right\}. \nonumber
\end{align*}
}
Thus, $\Tilde{\bV}_0|\text{rest}\sim\text{MVN}\left(\bC^{-1}\bb, \bC^{-1}\right)$ where $\bb = \frac{\bL_0-\bmm_{\delta_0}-\phi_0\bU_0}{\tau_{\psi_0}^2}+\frac{\gamma_v\left(\bL_1-\bmm_{\delta_1}-\Tilde{\bV}_1-\phi_1\bU_1\right)}{\tau_{\psi_1}^2}$ and $\bC=\frac{\mathbb{I}_G}{\tau_{\psi_0}^2}+\frac{\gamma_v^2\mathbb{I}_G}{\tau_{\psi_1}^2}+\frac{\bR^{-1}(\rho_v,\kappa_v)}{\sigma_{v,0}^2}$.

\vspace{10pt}
\noindent
{\bf Full conditional distribution for $\Tilde{\bV}_1$}

\noindent
Note that $\Tilde{\bV}_1\sim\text{MVN}\left\{\mathbf{0}_G, \sigma_{v,1}^2\bR(\rho_v,\kappa_v)\right\}$. Then,
\small{
\begin{align*}
    &q\left(\Tilde{\bV}_1|\text{rest}\right) \\
    &\propto \calL\left(\bL_1|\Tilde{\bV}_1, \text{rest}\right)\pi\left(\Tilde{\bV}_1\right) \\ 
    &\propto\exp\left\{-\frac{\left(\bL_1-\bmm_{\delta_1}-\Tilde{\bV}_1-\gamma_v\Tilde{\bV}_0-\phi_1\bU_1\right)^{\top}\left(\bL_1-\bmm_{\delta_1}-\Tilde{\bV}_1-\gamma_v\Tilde{\bV}_0-\phi_1\bU_1\right)}{2\tau_{\psi_1}^2}\right\} \times \\
    &\hspace{15pt}\exp\left\{-\frac{\Tilde{\bV}_1^{\top}\bR^{-1}(\rho_v,\kappa_v)\Tilde{\bV}_1}{2\sigma_{v,1}^2}\right\} \nonumber \\
    &\propto\exp\left\{-\frac{\Tilde{\bV}_1^{\top}\Tilde{\bV}_1-2\Tilde{\bV}_1^{\top}\left(\bL_1-\bmm_{\delta_1}-\gamma_v\Tilde{\bV}_0-\phi_1\bU_1\right)}{2\tau_{\psi_1}^2}\right\}\exp\left\{-\frac{\Tilde{\bV}_1^{\top}\bR^{-1}(\rho_v,\kappa_v)\Tilde{\bV}_1}{2\sigma_{v,1}^2}\right\} \\
    &\propto \exp\left[-\frac{1}{2}\left\{\Tilde{\bV}_1^{\top}\left(\frac{\mathbb{I}_G}{\tau_{\psi_1}^2}+\frac{\bR^{-1}(\rho_v,\kappa_v)}{\sigma_{v,1}^2}\right)\Tilde{\bV}_1-\frac{2\Tilde{\bV}_1^{\top}\left(\bL_1-\bmm_{\delta_1}-\gamma_v\Tilde{\bV}_0-\phi_1\bU_1\right)}{\tau_{\psi_1}^2}\right\}\right]
\end{align*}
}
Thus, $\Tilde{\bV}_1|\text{rest}\sim\text{MVN}\left(\bC^{-1}\bb, \bC^{-1}\right)$ where $\bb = \frac{\bL_1-\bmm_{\delta_1}-\gamma_v\Tilde{\bV}_0-\phi_1\bU_1}{\tau_{\psi_1}^2}$ and $\bC=\frac{\mathbb{I}_G}{\tau_{\psi_1}^2}+\frac{\bR^{-1}(\rho_v,\kappa_v)}{\sigma_{v,1}^2}$.

\vspace{10pt}
\noindent
{\bf Full conditional distribution for $\sigma_{u,a}^2$ and $\sigma_{v,a}^2$ for $a\in\left\{0,1\right\}$}

\noindent
Let the prior be $\sigma_{u,a}^2\sim\text{Inverse-Gamma}(a_u, b_u)$. Then,
\begin{align*}
    &p\left(\sigma_{u,a}^2|\text{rest}\right) \\
    &\propto \left(\frac{1}{\sigma_{u,a}^2}\right)^{G/2}\exp\left\{-\frac{\Tilde{\bU}_a^{\top}\bR^{-1}(\rho_u,\kappa_u)\Tilde{\bU}_a}{2\sigma_{u,a}^2}\right\}\times\left(\frac{1}{\sigma_{u,a}^2}\right)^{a_u+1}\exp\left(-\frac{b_u}{\sigma_{u,a}^2}\right) \\
    &\propto \left(\frac{1}{\sigma_{u,a}^2}\right)^{G/2+a_u+1}\exp\left\{-\frac{\Tilde{\bU}_a^{\top}\bR^{-1}(\rho_u,\kappa_u)\Tilde{\bU}_a/2+b_u}{\sigma_{u,a}^2}\right\}.
\end{align*}
Therefore, $\sigma_{u,a}^2|\text{rest}\sim\text{Inverse-Gamma}\left\{\frac{G}{2}+a_u,\frac{\Tilde{\bU}_a^{\top}\bR^{-1}(\rho_u,\kappa_u)\Tilde{\bU}_a}{2}+b_u \right\}$. A similar argument with using $\sigma_{v,a}^2\sim\text{Invese-Gamma}(a_v, b_v)$ produces a full conditional distribution 
\begin{equation*}
    \sigma_{v,a}^2|\text{rest}\sim\text{Inverse-Gamma}\left\{\frac{G}{2}+a_v,\frac{\Tilde{\bV}_a^{\top}\bR^{-1}(\rho_v,\kappa_v)\Tilde{\bV}_a}{2}+b_v \right\}.
\end{equation*}

\vspace{10pt}
\noindent
{\bf Full conditional distribution for $\tau_{\psi_0}^2$, $\tau_{\psi_1}^2$}. 

\noindent
Let the prior be $\tau_{\psi_a}^2\sim\text{Inverse-Gamma}(a_\psi,b_\psi)$. Then,
\small{\begin{align*}
    &p\left(\tau_{\psi_a}^2|\text{rest}\right) \\
    &\propto \left(\frac{1}{\tau_{\psi_a}^2}\right)^{G/2}\exp\left\{-\frac{\left(\bL_a-\bmm_{\delta_a}-\bV_a-\phi_a\bU_a\right)^{\top}\left(\bL_a-\bmm_{\delta_a}-\bV_a-\phi_a\bU_a\right)}{2\tau_{\psi_a}^2}\right\}\times\pi(\tau_{\psi_a}^2) \\
    &\propto\left(\frac{1}{\tau_{\psi_a}^2}\right)^{G/2+a_\psi+1}\exp\left\{-\frac{\left(\bL_a-\bmm_{\delta_a}-\bV_a-\phi_a\bU_a\right)^{\top}\left(\bL_a-\bmm_{\delta_a}-\bV_a-\phi_a\bU_a\right)/2+b_\psi}{\tau_{\psi_a}^2}\right\}.
\end{align*}}
Thus, $\tau_{\psi_a}^2|\text{rest}\sim\text{Inverse-Gamma}\left(\frac{G}{2}+a_\psi,\frac{\left(\bL_a-\bmm_{\delta_a}-\bV_a-\phi_a\bU_a\right)^{\top}\left(\bL_a-\bmm_{\delta_a}-\bV_a-\phi_a\bU_a\right)}{2}+b_\psi\right)$.

\vspace{10pt}
\noindent
{\bf Full conditional distribution for $\tau_{0}^2$, $\tau_{1}^2$}

\noindent
Let the prior be $\tau_a^2\sim\text{Inverse-Gamma}(a_y,b_y)$. Then,
\small{
\begin{align*}
    &p\left(\tau_{a}^2|\text{rest}\right) \\
    &\propto \left(\frac{1}{\tau_a^2}\right)^{n/2}\exp\left\{-\frac{\left(\bY_a-\bmm_{\beta_a}-\bH\bU_a\right)^{\top}\left(\bY_a-\bmm_{\beta_a}-\bH\bU_a\right)}{2\tau_a^2}\right\}\times\left(\frac{1}{\tau_a^2}\right)^{a_y+1}\exp\left(-\frac{b_y}{\tau_a^2}\right) \\
    &\propto \left(\frac{1}{\tau_a^2}\right)^{n/2+a_y+1}\exp\left\{-\frac{\left(\bY_a-\bmm_{\beta_a}-\bH\bU_a\right)^{\top}\left(\bY_a-\bmm_{\beta_a}-\bH\bU_a\right)/2+b_y}{\tau_a^2}\right\}.
\end{align*}
}
Thus, $\tau_a^2|\text{rest}\sim\text{Inverse-Gamma}\left(\frac{n}{2}+a_y, \frac{\left(\bY_a-\bmm_{\beta_a}-\bH\bU_a\right)^{\top}\left(\bY_a-\bmm_{\beta_a}-\bH\bU_a\right)}{2}+b_y\right)$.

\subsection{Data sources of the covariates}\label{supp::source}

Original data do not contain grid-level covariates, since grid cells are created in the analysis stage, not in the data-collection stage. Therefore, to keep site-level and grid-level covariates on the same scale, simultaneously at the coordinates of each survey site and each centroid of the grid cell, we extract the values of the aforementioned covariates from various data sources. Distances to the shoreline are calculated by the great circle distance between survey points or grid centroids and the seashore. The coordinates for seashore are available using the high resolution shoreline layer of the Global Self-consistent, Hierarchical, High-resolution Geography (GSHHG) Version 2.3.7 \citep{wessel1996global}. For depth values, they are extracted using AusBathyTopo (Australia) 250 m 2023 - A High-resolution Depth Model \citep{Beaman2023}. For the mean level of chlorophyll a and the minimum sea surface temperature, the values are extracted from BIO-ORACLE \citep{tyberghein2012bio}, which represent the monthly average values for the years 2002-2009. Human population within 100 km radius of each survey point and grid cell is calculated by using the Socioeconomic Data and Application Centre (SEDAC) Gridded Population Of The World database for the year 2000 \citepalias{ciesin_2018}. Distances to the market are the great circle distance between the survey points or grid centroids and the nearest provincial or national capital, where the coordinates are identified from the World Cities base map layer \citep{esri_2023}. The habitat types at the survey sites are brought from the original MPA data, while the habitat types at the grid cells are matched with those of the nearest survey sites. 


\subsection{Sensitivity to the resolution of grid cells}\label{supp::grid_sensitivity}

To check the sensitivity of the marginal posterior distributions of $\triangle$ according to the grid cell resolution, we fit the proposed model with coarser and finer resolutions. For coarser resolutions, we cover the spatial domain with grid cells $1.0\times1.0\text{ degree}^2$, while we use grid cells $0.6\times0.6\text{ degree}^2$ for finer resolution. The number of grid cells with the coarser resolution is $G=293$, while that with the finer resolution is $G=813$. We use the same prior distributions as described in Section 5.2, except that we assign $\log(\rho_u)$ and $\log(\rho_v)$ with $\text{Normal}\{\log(400), 0.5\}$ for the coarser resolution. 
The posterior mean of $\triangle$ with the resolution used in the main manuscript is $0.71$ with the 95\% credible interval being $(-0.24, 1.71)$. Similarly, those with coarser and finer resolutions are $0.343$ and $0.960$, respectively, while the 95\% credible intervals are $(-0.858, 1.652)$ and $(-0.358, 2.191)$ respectively.

\subsection{Overlap in propensity scores between the policy groups}\label{supp::ps_overlap}

\begin{figure}[t]
    \centering
    \includegraphics[scale=0.5]{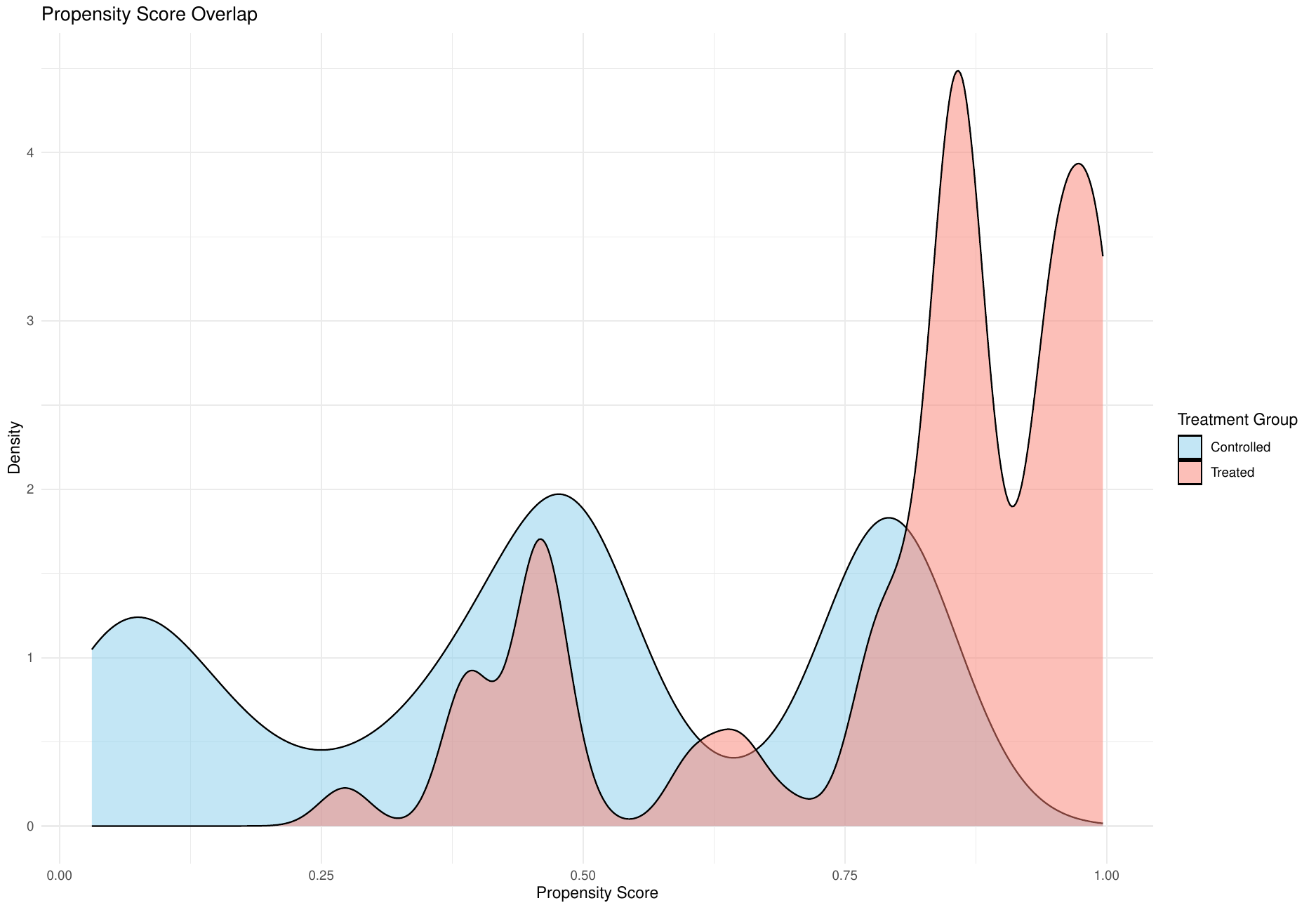}
    \caption{Overlap check for the propensity scores between the policy groups.}    
    \label{fig::ps_overlap}
\end{figure}

From the model specification, the propensity scores are estimated in a grid level as $\frac{\lambda_{1,g}}{\lambda_{0,g}+\lambda_{1,g}}$ for $g=1,\cdots,G$. We then define the site-level propensity scores as those at the grid cells in which the sites are located. Figure~\ref{fig::ps_overlap} represents the estimated densities of the site-level propensity scores for each policy group. We observe from the plot that two groups are relatively imbalanced for the propensity scores higher than $0.9$ and lower than $0.2$. We conjecture that the imbalances occur due to how the propensity scores are defined by the model specification as noted above. Furthermore, we speculate that this arises from the proposed method postulating only the outcome regression models, not the propensity score models. This comes as one of the drawbacks of the proposed models, requiring future work to address this problem.  

\begin{singlespace}
	\bibliographystyle{apalike}
	\bibliography{refs}
\end{singlespace}